\documentclass{llncs}

\usepackage{etex}
\usepackage{llncsdoc}
\usepackage[usenames,dvipsnames]{pstricks}
\usepackage{epsfig}
\usepackage{palatino}
\RequirePackage{ifthen}
\usepackage{latexsym}
\RequirePackage{amsmath}
\RequirePackage{amssymb}
\RequirePackage{xspace}
\RequirePackage{graphics}
\usepackage{xcolor}
\usepackage{wrapfig}
\usepackage{keyval}
\usepackage{xspace}
\usepackage{mathrsfs}
\usepackage{graphicx}
\usepackage{paralist}
\usepackage{float}
\usepackage{mathtools}
\usepackage{longtable}

\usepackage[font=footnotesize]{caption}
\DeclareCaptionType{copyrightbox}
\usepackage{subcaption}
\usepackage{booktabs}
\usepackage{arydshln}

\usepackage{amsmath,amssymb,url,listings,mathrsfs}
\usepackage{framed}
\usepackage{lipsum}

\usepackage{tikz}
\usetikzlibrary{calc}
\usetikzlibrary{matrix}

\definecolor{reddish}{rgb}{1,.8,0.8}
\definecolor{blueish}{rgb}{0.8,.8,1}
\definecolor{greenish}{rgb}{.8,1,0.8}
\definecolor{yellowish}{rgb}{1,1,.20}

\definecolor{webred}{rgb}{0.5,0,0}
\definecolor{webblue}{rgb}{0,0,0.8}
\usepackage[pdftex,colorlinks,citecolor=webblue,linkcolor=webred,]{hyperref}

\newcommand{\intervals}{\relax \ifmmode \mathbb{IR} \else $\mathbb{IR}$ \fi}
\newcommand{\nnintervals}{\relax \ifmmode \mathbb{IR}^{+} \else $\mathbb{IR}^{+}$ \fi}

\newcommand{\seman}[1]{\relax \ifmmode [\![#1]\!] \else $[\![#1]\!]$ \fi}

\newcommand{\tbundle}{\texttt{TransformBundle} }

\newcommand{\num}[1]{\relax\ifmmode \mathbb #1\else $\mathbb #1$\fi}
\newcommand{\nnnum}[1]{\relax\ifmmode 
  {\mathbb #1}_{\geq 0} \else ${\mathbb #1}_{\geq 0}$
  \fi}
\newcommand{\npnum}[1]{\relax\ifmmode 
  {\mathbb #1}_{\leq 0} \else ${\mathbb #1}_{\leq 0}$
  \fi}
\newcommand{\pnum}[1]{\relax\ifmmode 
  {\mathbb #1}_{> 0} \else ${\mathbb #1}_{> 0}$
  \fi}
\newcommand{\nnum}[1]{\relax\ifmmode 
  {\mathbb #1}_{< 0} \else ${\mathbb #1}_{< 0}$
  \fi}
\newcommand{\plnum}[1]{\relax\ifmmode 
  {\mathbb #1}_{+} \else ${\mathbb #1}_{+}$
  \fi}
\newcommand{\nenum}[1]{\relax\ifmmode 
  {\mathbb #1}_{-} \else ${\mathbb #1}_{-}$
  \fi}

\newcommand{\reals}{{\num R}}                    %reals
\newcommand{\extb}[1]{\relax\ifmmode {\sf ExtBeh}_{#1} \else ${\sf ExtBeh}_{#1}$\fi} 
\newcommand{\tdists}[1]{\relax\ifmmode {\sf Tdists}_{#1} \else ${\sf Tdists}_{#1}$\fi} 

\newcommand{\exec}[1]{\relax\ifmmode {\sf Execs}_{#1} \else ${\sf Exec}_{#1}$\fi} 
\newcommand{\execf}[1]{\relax\ifmmode {\sf Execs}^*_{#1} \else ${\sf Exec}^*_{#1}$\fi} 
\newcommand{\execi}[1]{\relax\ifmmode {\sf Execs}^\omega_{#1} \else ${\sf Exec}^\omega_{#1}$\fi} 

\newcommand{\ctrace}[1]{\relax\ifmmode {\sf Ctraces}_{#1} \else ${\sf Ctraces}_{#1}$\fi} 

\newcommand{\trace}[1]{\relax\ifmmode {\sf Traces}_{#1} \else ${\sf Traces}_{#1}$\fi} 
\newcommand{\tracef}[1]{\relax\ifmmode {\sf Traces}^*_{#1} \else ${\sf Traces}^*_{#1}$\fi} 
\newcommand{\tracei}[1]{\relax\ifmmode {\sf Traces}^\omega_{#1} \else ${\sf Traces}^\omega_{#1}$\fi} 

\newcommand{\frag}[1]{\relax\ifmmode {\sf Frags}_{#1} \else ${\sf Frags}_{#1}$\fi} 
\newcommand{\fragf}[1]{\relax\ifmmode {\sf Frags}^*_{#1} \else ${\sf Frags}^*_{#1}$\fi} 
\newcommand{\fragi}[1]{\relax\ifmmode {\sf Frags}^\omega_{#1} \else ${\sf Frags}^\omega_{#1}$\fi} 

\newcommand{\reach}[1]{\relax\ifmmode {\sf Reach}_{#1} \else ${\sf Reach}_{#1}$\fi} 
\newcommand{\pair}[2]{\relax\ifmmode \langle #1, #2 \rangle \else $\langle #1, #2 \rangle$\fi} 

\newcommand{\TE}{\relax\ifmmode \mathit{Time} \else $\mathit{Time}$ \fi} 
\newcommand{\EQ}{\relax\ifmmode \mathit{Enq} \else $\mathit{Enq}$ \fi} 
\newcommand{\DQ}{\relax\ifmmode \mathit{Deq} \else $\mathit{DeqTime}$ \fi} 
\newcommand{\E}{\relax\ifmmode \mathsf{E} \else $\mathsf{E}$ \fi}

\newcommand{\loc}{\relax\ifmmode \mathit{loc} \else $\mathit{loc}$ \fi}
\newcommand{\abs}{\relax\ifmmode \mathit{abs} \else $\mathit{abs}$ \fi}

\newcommand{\pf}{\par\noindent{\bf Proof:}~}

\def\A{{\cal A}} % HA
\def\E{{\cal E}} % HA
\def\I{{\cal I}} % environment sequence
\def\R{{\cal R}} % relation
\def\T{{\cal T}} % set of trajectories
\def\U{{\cal U}} % set of trajectories
\newcommand{\col}[1]{\relax\ifmmode \mathscr #1\else $\mathscr #1$\fi}

\definecolor{HIOAcolor}{rgb}{0.776,0.22,0.07}

\newcommand{\SC}[2]{\relax\ifmmode {\tt Scount}(#1,#2) \else ${\tt Scount}(#1,#2)$\fi} 
\newcommand{\SCM}[2]{\relax\ifmmode {\tt Smin}(#1,#2) \else ${\tt Smin}(#1,#2)$\fi} 
\newcommand{\Aut}[1]{\relax\ifmmode {\tt Aut}(#1) \else ${\tt Aut}(#1)$\fi}

\newcommand{\For}{{\bf for}}

\newcommand{\remove}[1]{}
\newcommand{\salg}[1]{\relax\ifmmode {\mathcal F}_{#1}\else ${\mathcal F}_{#1}$\fi} 
\newcommand{\msp}[1]{\relax\ifmmode (#1, \salg{#1}) \else $(#1, \salg{#1})$\fi} 
\newcommand{\msprod}[2]{\relax\ifmmode ( #1 \times #2, \salg{#1} \otimes \salg{#2}) \else $(#1 \times #2, \salg{#1} \otimes \salg{#2})$\fi} 
\newcommand{\dist}[1]{\relax\ifmmode {\mathcal P}\msp{#1}
  \else ${\mathcal P}\msp{#1}$\fi} 
\newcommand{\subdist}[1]{\relax\ifmmode {\mathcal S}{\mathcal P}\msp{#1} 
  \else ${\mathcal S}{\mathcal P}\msp{#1}$\fi} 
\newcommand{\disc}[1]{\relax\ifmmode {\sf Disc}(#1)
  \else ${\sf Disc}(#1)$\fi} 

\newcommand{\Trajeq}{\relax\ifmmode {\mathcal R}_\T \else ${\mathcal R}_\T$\fi} 
\newcommand{\Acteq}{\relax\ifmmode {\mathcal R}_A \else ${\mathcal R}_A$\fi} 
\newcommand{\noop}{\relax\ifmmode \lambda \else $\lambda$\fi} 
\newcommand{\close}[1]{\relax\ifmmode \overline{#1} \else $\overline{#1}$\fi}

\newcommand{\tup}[1]
           {
             \relax\ifmmode
             \langle #1 \rangle
             \else $\langle$ #1 $\rangle$ \fi
           }

\newcommand{\lit}[1]{ \relax\ifmmode
                \mathord{\mathcode`\-="702D\sf #1\mathcode`\-="2200}
                \else {\it #1} \fi }

\newcommand{\figuresize}{\scriptsize}

\lstdefinelanguage{ioa}{
  basicstyle=\figuresize,
  keywordstyle=\bf \figuresize,
  identifierstyle=\it \figuresize,
  emphstyle=\tt \figuresize,
  mathescape=true,
  tabsize=20,
  sensitive=false,
  columns=fullflexible,
  keepspaces=false,
  flexiblecolumns=true,
  basewidth=0.05em,
  moredelim=[il][\rm]{//},
  moredelim=[is][\sf \figuresize]{!}{!},
  moredelim=[is][\bf \figuresize]{*}{*},
  keywords={automaton,and, 
         choose,const,continue, components,
         discrete, do, derived,
         eff, external,else, elseif, evolve, end,each,
         fi,for, forward, from,
         hidden,
         in,input,internal,if,invariant, initially, imports,
     let,
     or, output, operators, od, of,
     pre, prob,
     return,
     such,satisfies, stop, signature, simulation, state, stochastic,
     trajectories,trajdef, transitions, that,then, type, types, to, tasks,
     variables, vocabulary, uni,
     when,where, with,while},
  emph={set, seq, tuple, map, array, enumeration},   
   literate=
        {(}{{$($}}1
        {)}{{$)$}}1
        {\\in}{{$\in\ $}}1
        {\\preceq}{{$\preceq\ $}}1
        {\\subset}{{$\subset\ $}}1
        {\\subseteq}{{$\subseteq\ $}}1
        {\\supset}{{$\supset\ $}}1
        {\\supseteq}{{$\supseteq\ $}}1
        {\\forall}{{$\forall$}}1
        {\\le}{{$\le\ $}}1
        {\\ge}{{$\ge\ $}}1
        {\\gets}{{$\gets\ $}}1
        {\\cup}{{$\cup\ $}}1
        {\\cap}{{$\cap\ $}}1
        {\\langle}{{$\langle$}}1
        {\\rangle}{{$\rangle$}}1
        {\\exists}{{$\exists\ $}}1
        {\\bot}{{$\bot$}}1
        {\\rip}{{$\rip$}}1
        {\\emptyset}{{$\emptyset$}}1
        {\\notin}{{$\notin\ $}}1
        {\\not\\exists}{{$\not\exists\ $}}1
        {\\ne}{{$\ne\ $}}1
        {\\to}{{$\to\ $}}1
        {\\implies}{{$\implies\ $}}1
        {<}{{$<\ $}}1
        {>}{{$>\ $}}1
        {=}{{$=\ $}}1
        {~}{{$\neg\ $}}1
        {|}{{$\mid$}}1
        {'}{{$^\prime$}}1
        {\\A}{{$\forall\ $}}1
        {\\E}{{$\exists\ $}}1
        {\\/}{{$\vee\,$}}1
        {\\vee}{{$\vee\,$}}1
        {/\\}{{$\wedge\,$}}1
        {\\wedge}{{$\wedge\,$}}1
        {=>}{{$\Rightarrow\ $}}1
        {->}{{$\rightarrow\ $}}1
        {<=}{{$\Leftarrow\ $}}1
        {<-}{{$\leftarrow\ $}}1
        {~=}{{$\neq\ $}}1
        {\\U}{{$\cup\ $}}1
        {\\I}{{$\cap\ $}}1
        {|-}{{$\vdash\ $}}1
        {-|}{{$\dashv\ $}}1
        {<<}{{$\ll\ $}}2
        {>>}{{$\gg\ $}}2
        {||}{{$\|$}}1
        {[}{{$[$}}1
        {]}{{$\,]$}}1
        {[[}{{$\langle$}}1
        {]]]}{{$]\rangle$}}1
        {]]}{{$\rangle$}}1
        {<=>}{{$\Leftrightarrow\ $}}2
        {<->}{{$\leftrightarrow\ $}}2
        {(+)}{{$\oplus\ $}}1
        {(-)}{{$\ominus\ $}}1
        {_i}{{$_{i}$}}1
        {_j}{{$_{j}$}}1
        {_{i,j}}{{$_{i,j}$}}3
        {_{j,i}}{{$_{j,i}$}}3
        {_0}{{$_0$}}1
        {_1}{{$_1$}}1
        {_2}{{$_2$}}1
        {_n}{{$_n$}}1
        {_p}{{$_p$}}1
        {_k}{{$_n$}}1
        {-}{{$\ms{-}$}}1
        {@}{{}}0
        {\\delta}{{$\delta$}}1
        {\\R}{{$\R$}}1
        {\\Rplus}{{$\Rplus$}}1
        {\\N}{{$\N$}}1
        {\\times}{{$\times\ $}}1
        {\\tau}{{$\tau$}}1
        {\\alpha}{{$\alpha$}}1
        {\\beta}{{$\beta$}}1
        {\\gamma}{{$\gamma$}}1
        {\\ell}{{$\ell\ $}}1
        {--}{{$-\ $}}1
        {\\TT}{{\hspace{1.5em}}}3        
      }

\lstdefinelanguage{ioaNums}[]{ioa}
{
  numbers=left,
  numberstyle=\tiny,
  stepnumber=2,
  numbersep=4pt
}

\lstdefinelanguage{ioaNumsRight}[]{ioa}
{
  numbers=right,
  numberstyle=\tiny,
  stepnumber=2,
  numbersep=4pt
}

\lstnewenvironment{IOA}%
  {\lstset{language=IOA}}
  {}

\lstnewenvironment{IOANums}%
  {
  \if@firstcolumn
    \lstset{language=IOA, numbers=left, firstnumber=auto}
  \else
    \lstset{language=IOA, numbers=right, firstnumber=auto}
  \fi
  }
  {}

\lstnewenvironment{IOANumsRight}%
  {
    \lstset{language=IOA, numbers=right, firstnumber=auto}
  }
  {}

\newcommand{\linefigioa}[9]{

}

\lstdefinelanguage{ioaLang}{%
  basicstyle=\ttfamily\small,
  keywordstyle=\rmfamily\bfseries\small,
  identifierstyle=\small,
  keywords={assumes,automaton,axioms,backward,bounds,by,case,choose,components,const,d,det,discrete,do,eff,else,elseif,ensuring,enumeration,evolve,fi,fire,follow,for,forward,from,hidden,if,in,%
    input,initially,internal,invariant,let, local,od,of,output,pre,schedule,signature,so,%
    simulation,states,state,variables, tasks, stop,tasks,that,then,to,trajdef,trajectory,trajectories,transitions,tuple,type,
    uniform,union,urgent,uses,when,where,while,yield},
  literate=
        {\\in}{{$\in$}}1
        {\\preceq}{{$\preceq$}}1
        {\\subset}{{$\subset$}}1
        {\\subseteq}{{$\subseteq$}}1
        {\\supset}{{$\supset$}}1
        {\\supseteq}{{$\supseteq$}}1
        {\\rho}{{$\rho$}}1
        {\\infty}{{$\infty$}}1
        {<}{{$<$}}1
        {>}{{$>$}}1
        {=}{{$=$}}1
        {~}{{$\neg$}}1 
        {|}{{$\mid$}}1
        {'}{{$^\prime$}}1
        {\\A}{{$\forall$}}1 {\\E}{{$\exists$}}1
        {\\/}{{$\vee$}}1 {/\\}{{$\wedge$}}1 
        {=>}{{$\Rightarrow$}}1 
        {->}{{$\rightarrow$}}1 
        {<=}{{$\leq$}}1 {>=}{{$\geq$}}1 {~=}{{$\neq$}}1
        {\\U}{{$\cup$}}1 {\\I}{{$\cap$}}1
        {|-}{{$\vdash$}}1 {-|}{{$\dashv$}}1
        {<<}{{$\ll$}}2 {>>}{{$\gg$}}2
        {||}{{$\|$}}1
        {<=>}{{$\Leftrightarrow$}}2 
        {<->}{{$\leftrightarrow$}}2
        {(+)}{{$\oplus$}}1
        {(-)}{{$\ominus$}}1
}

\lstdefinelanguage{bigIOALang}{%
  basicstyle=\ttfamily,
  keywordstyle=\rmfamily\bfseries,
  identifierstyle=,
  keywords={assumes,automaton,axioms,backward,by,case,choose,components,const,%
    d,det,discrete,do,eff,else,elseif,ensuring,enumeration,evolve,fi,for,forward,from,hidden,if,in%
    input,initially,internal,invariant,local,od,of,output,pre,schedule,signature,so,%
    tasks, simulation,states,stop,tasks,that,then,to,trajdef,trajectories,transitions,tuple,type,union,urgent,uses,when,where,yield},
  literate=
        {\\in}{{$\in$}}1
        {\\preceq}{{$\preceq$}}1
        {\\subset}{{$\subset$}}1
        {\\subseteq}{{$\subseteq$}}1
        {\\supset}{{$\supset$}}1
        {\\supseteq}{{$\supseteq$}}1
        {<}{{$<$}}1
        {>}{{$>$}}1
        {=}{{$=$}}1
        {~}{{$\neg$}}1 
        {|}{{$\mid$}}1
        {'}{{$^\prime$}}1
        {\\A}{{$\forall$}}1 {\\E}{{$\exists$}}1
        {\\/}{{$\vee$}}1 {/\\}{{$\wedge$}}1 
        {=>}{{$\Rightarrow$}}1 
        {->}{{$\rightarrow$}}1 
        {<=}{{$\leq$}}1 {>=}{{$\geq$}}1 {~=}{{$\neq$}}1
        {\\U}{{$\cup$}}1 {\\I}{{$\cap$}}1
        {|-}{{$\vdash$}}1 {-|}{{$\dashv$}}1
        {<<}{{$\ll$}}2 {>>}{{$\gg$}}2
        {||}{{$\|$}}1
        {<=>}{{$\Leftrightarrow$}}2 
        {<->}{{$\leftrightarrow$}}2
        {(+)}{{$\oplus$}}1
        {(-)}{{$\ominus$}}1
}

\lstnewenvironment{BigIOA}%
  {\lstset{language=bigIOALang,basicstyle=\ttfamily}
   \csname lst@SetFirstLabel\endcsname}
  {\csname lst@SaveFirstLabel\endcsname\vspace{-4pt}\noindent}

\lstnewenvironment{SmallIOA}%
  {\lstset{language=ioaLang,basicstyle=\ttfamily\scriptsize}
   \csname lst@SetFirstLabel\endcsname}
  {\csname lst@SaveFirstLabel\endcsname\noindent}

\newcommand{\true}{\relax\ifmmode \mathit true \else \em true \/\fi}
\newcommand{\false}{\relax\ifmmode \mathit false \else \em false \/\fi}

\newlength{\bracklen}

\newcommand{\tri}[3]{\ensuremath{\mathit{#1}^\mathit{#2}_\mathit{#3}}}

\newcommand{\sugLocalVars}[2]{\ifthenelse{\equal{}{#2}}%
                             {\tri{localVars}{#1}{desug}}%
                             {\tri{localVars}{#1}{#2,desug}}}
\newcommand{\sugVars}[2]{\ifthenelse{\equal{}{#2}}%
                        {\tri{vars}{#1}{desug}}%
                        {\tri{vars}{#1}{#2,desug}}}

\newenvironment{subSyntax}{\begin{array}{l}}{\end{array}}

\newcommand{\ms}[1]{\ifmmode%
\mathord{\mathcode`-="702D\it #1\mathcode`\-="2200}\else%
$\mathord{\mathcode`-="702D\it #1\mathcode`\-="2200}$\fi}

\def\A{{\cal A}} % TA
\def\T{{\cal T}} % set of trajectories

\lstdefinelanguage{pvs}{
  basicstyle=\tt \figuresize,
  keywordstyle=\sc \figuresize,
  identifierstyle=\it \figuresize,
  emphstyle=\tt \figuresize,
  mathescape=true,
  tabsize=20,
  sensitive=false,
  columns=fullflexible,
  keepspaces=false,
  flexiblecolumns=true,
  basewidth=0.05em,
  moredelim=[il][\rm]{//},
  moredelim=[is][\sf \figuresize]{!}{!},
  moredelim=[is][\bf \figuresize]{*}{*},
  keywords={and, 
         begin,
         cases, const,
         do,
         external, else, exists, end, endcases, endif,
         fi,for, forall, from,
         hidden,
         in, if, importing,
     let, lambda, lemma,
     measure, 
     not,
     or, of,
     return, recursive,
     stop, 
     theory, that,then, type, types, type+, to, theorem,
     var,
     with,while},
  emph={nat, setof, sequence, eq, tuple, map, array, enumeration, bool, real, exp, nnreal, posreal},   
   literate=
        {(}{{$($}}1
        {)}{{$)$}}1
        {\\in}{{$\in\ $}}1
        {\\mapsto}{{$\rightarrow\ $}}1
        {\\preceq}{{$\preceq\ $}}1
        {\\subset}{{$\subset\ $}}1
        {\\subseteq}{{$\subseteq\ $}}1
        {\\supset}{{$\supset\ $}}1
        {\\supseteq}{{$\supseteq\ $}}1
        {\\forall}{{$\forall$}}1
        {\\le}{{$\le\ $}}1
        {\\ge}{{$\ge\ $}}1
        {\\gets}{{$\gets\ $}}1
        {\\cup}{{$\cup\ $}}1
        {\\cap}{{$\cap\ $}}1
        {\\langle}{{$\langle$}}1
        {\\rangle}{{$\rangle$}}1
        {\\exists}{{$\exists\ $}}1
        {\\bot}{{$\bot$}}1
        {\\rip}{{$\rip$}}1
        {\\emptyset}{{$\emptyset$}}1
        {\\notin}{{$\notin\ $}}1
        {\\not\\exists}{{$\not\exists\ $}}1
        {\\ne}{{$\ne\ $}}1
        {\\to}{{$\to\ $}}1
        {\\implies}{{$\implies\ $}}1
        {<}{{$<\ $}}1
        {>}{{$>\ $}}1
        {=}{{$=\ $}}1
        {~}{{$\neg\ $}}1
        {|}{{$\mid$}}1
        {'}{{$^\prime$}}1
        {\\A}{{$\forall\ $}}1
        {\\E}{{$\exists\ $}}1
        {\\/}{{$\vee\,$}}1
        {\\vee}{{$\vee\,$}}1
        {/\\}{{$\wedge\,$}}1
        {\\wedge}{{$\wedge\,$}}1
        {->}{{$\rightarrow\ $}}1
        {=>}{{$\Rightarrow\ $}}1
        {->}{{$\rightarrow\ $}}1
        {<=}{{$\Leftarrow\ $}}1
        {<-}{{$\leftarrow\ $}}1
        {~=}{{$\neq\ $}}1
        {\\U}{{$\cup\ $}}1
        {\\I}{{$\cap\ $}}1
        {|-}{{$\vdash\ $}}1
        {-|}{{$\dashv\ $}}1
        {<<}{{$\ll\ $}}2
        {>>}{{$\gg\ $}}2
        {||}{{$\|$}}1
        {[}{{$[$}}1
        {]}{{$\,]$}}1
        {[[}{{$\langle$}}1
        {]]]}{{$]\rangle$}}1
        {]]}{{$\rangle$}}1
        {<=>}{{$\Leftrightarrow\ $}}2
        {<->}{{$\leftrightarrow\ $}}2
        {(+)}{{$\oplus\ $}}1
        {(-)}{{$\ominus\ $}}1
        {_i}{{$_{i}$}}1
        {_j}{{$_{j}$}}1
        {_{i,j}}{{$_{i,j}$}}3
        {_{j,i}}{{$_{j,i}$}}3
        {_0}{{$_0$}}1
        {_1}{{$_1$}}1
        {_2}{{$_2$}}1
        {_n}{{$_n$}}1
        {_p}{{$_p$}}1
        {_k}{{$_n$}}1
        {-}{{$\ms{-}$}}1
        {@}{{}}0
        {\\delta}{{$\delta$}}1
        {\\R}{{$\R$}}1
        {\\Rplus}{{$\Rplus$}}1
        {\\N}{{$\N$}}1
        {\\times}{{$\times\ $}}1
        {\\tau}{{$\tau$}}1
        {\\alpha}{{$\alpha$}}1
        {\\beta}{{$\beta$}}1
        {\\gamma}{{$\gamma$}}1
        {\\ell}{{$\ell\ $}}1
        {--}{{$-\ $}}1
        {\\TT}{{\hspace{1.5em}}}3        
      }

\lstdefinelanguage{BigPVS}{
  basicstyle=\tt,
  keywordstyle=\sc,
  identifierstyle=\it,
  emphstyle=\tt ,
  mathescape=true,
  tabsize=20,
  sensitive=false,
  columns=fullflexible,
  keepspaces=false,
  flexiblecolumns=true,
  basewidth=0.05em,
  moredelim=[il][\rm]{//},
  moredelim=[is][\sf \figuresize]{!}{!},
  moredelim=[is][\bf \figuresize]{*}{*},
  keywords={and, 
         begin,
         cases, const,
         do, datatype,
         external, else, exists, end, endif, endcases,
         fi,for, forall, from,
         hidden,
         in, if, importing,
     let, lambda, lemma,
     measure,
     not,
     or, of,
     return, recursive,
     stop, 
     theory, that,then, type, types, type+, to, theorem,
     var,
     with,while},
  emph={nat, setof, sequence, eq, tuple, map, array, first, rest, add, enumeration, bool, real, posreal, nnreal},   
   literate=
        {(}{{$($}}1
        {)}{{$)$}}1
        {\\in}{{$\in\ $}}1
        {\\mapsto}{{$\rightarrow\ $}}1
        {\\preceq}{{$\preceq\ $}}1
        {\\subset}{{$\subset\ $}}1
        {\\subseteq}{{$\subseteq\ $}}1
        {\\supset}{{$\supset\ $}}1
        {\\supseteq}{{$\supseteq\ $}}1
        {\\forall}{{$\forall$}}1
        {\\le}{{$\le\ $}}1
        {\\ge}{{$\ge\ $}}1
        {\\gets}{{$\gets\ $}}1
        {\\cup}{{$\cup\ $}}1
        {\\cap}{{$\cap\ $}}1
        {\\langle}{{$\langle$}}1
        {\\rangle}{{$\rangle$}}1
        {\\exists}{{$\exists\ $}}1
        {\\bot}{{$\bot$}}1
        {\\rip}{{$\rip$}}1
        {\\emptyset}{{$\emptyset$}}1
        {\\notin}{{$\notin\ $}}1
        {\\not\\exists}{{$\not\exists\ $}}1
        {\\ne}{{$\ne\ $}}1
        {\\to}{{$\to\ $}}1
        {\\implies}{{$\implies\ $}}1
        {<}{{$<\ $}}1
        {>}{{$>\ $}}1
        {=}{{$=\ $}}1
        {~}{{$\neg\ $}}1
        {|}{{$\mid$}}1
        {'}{{$^\prime$}}1
        {\\A}{{$\forall\ $}}1
        {\\E}{{$\exists\ $}}1
        {\\/}{{$\vee\,$}}1
        {\\vee}{{$\vee\,$}}1
        {/\\}{{$\wedge\,$}}1
        {\\wedge}{{$\wedge\,$}}1
        {->}{{$\rightarrow\ $}}1
        {=>}{{$\Rightarrow\ $}}1
        {->}{{$\rightarrow\ $}}1
        {<=}{{$\Leftarrow\ $}}1
        {<-}{{$\leftarrow\ $}}1
        {~=}{{$\neq\ $}}1
        {\\U}{{$\cup\ $}}1
        {\\I}{{$\cap\ $}}1
        {|-}{{$\vdash\ $}}1
        {-|}{{$\dashv\ $}}1
        {<<}{{$\ll\ $}}2
        {>>}{{$\gg\ $}}2
        {||}{{$\|$}}1
        {[}{{$[$}}1
        {]}{{$\,]$}}1
        {[[}{{$\langle$}}1
        {]]]}{{$]\rangle$}}1
        {]]}{{$\rangle$}}1
        {<=>}{{$\Leftrightarrow\ $}}2
        {<->}{{$\leftrightarrow\ $}}2
        {(+)}{{$\oplus\ $}}1
        {(-)}{{$\ominus\ $}}1
        {_i}{{$_{i}$}}1
        {_j}{{$_{j}$}}1
        {_{i,j}}{{$_{i,j}$}}3
        {_{j,i}}{{$_{j,i}$}}3
        {_0}{{$_0$}}1
        {_1}{{$_1$}}1
        {_2}{{$_2$}}1
        {_n}{{$_n$}}1
        {_p}{{$_p$}}1
        {_k}{{$_n$}}1
        {-}{{$\ms{-}$}}1
        {@}{{}}0
        {\\delta}{{$\delta$}}1
        {\\R}{{$\R$}}1
        {\\Rplus}{{$\Rplus$}}1
        {\\N}{{$\N$}}1
        {\\times}{{$\times\ $}}1
        {\\tau}{{$\tau$}}1
        {\\alpha}{{$\alpha$}}1
        {\\beta}{{$\beta$}}1
        {\\gamma}{{$\gamma$}}1
        {\\ell}{{$\ell\ $}}1
        {--}{{$-\ $}}1
        {\\TT}{{\hspace{1.5em}}}3        
      }

\lstdefinelanguage{pvsNums}[]{pvs}
{
  numbers=left,
  numberstyle=\tiny,
  stepnumber=2,
  numbersep=4pt
}

\lstdefinelanguage{pvsNumsRight}[]{pvs}
{
  numbers=right,
  numberstyle=\tiny,
  stepnumber=2,
  numbersep=4pt
}

\lstnewenvironment{BigPVS}%
  {\lstset{language=BigPVS}}
  {}

\lstnewenvironment{PVSNums}%
  {
  \if@firstcolumn
    \lstset{language=pvs, numbers=left, firstnumber=auto}
  \else
    \lstset{language=pvs, numbers=right, firstnumber=auto}
  \fi
  }
  {}

\lstnewenvironment{PVSNumsRight}%
  {
    \lstset{language=pvs, numbers=right, firstnumber=auto}
  }
  {}

\newcommand{\linefigpvs}[9]{

}

\lstdefinelanguage{pvsproof}{
  basicstyle=\tt \figuresize,
  mathescape=true,
  tabsize=4,
  sensitive=false,
  columns=fullflexible,
  keepspaces=false,
  flexiblecolumns=true,
  basewidth=0.05em,
}
\usepackage[boxed,linesnumbered,lined,commentsnumbered]{algorithm2e}

\title{Automatic Dynamic Parallelotope Bundles for Reachability Analysis of Nonlinear Systems}
\author{Edward Kim\inst{1} \and Stanley Bak \inst{2} \and Parasara Sridhar Duggirala \inst{1}}
\institute{University of North Carolina at Chapel Hill \and Stony Brook University}
\begin{document}

\maketitle

\begin{abstract}
  Reachable set computation is an important technique for the verification of safety properties of dynamical systems.
%
% one method proposed for solving this problem, based on parallelotope bundle reachability for
%
In this paper, we investigate reachable set computation for discrete nonlinear systems based on parallelotope bundles.
The algorithm relies on computing an upper bound on the supremum of a nonlinear function over a rectangular domain, which has been traditionally done using Bernstein polynomials.
%or nonlinear optimization tools.
 We strive to remove the manual step of parallelotope template selection to make the method fully automatic.
Furthermore, we show that changing templates dynamically during computations cans improve accuracy.
%
% One of the simplest algorithms for computing reachable sets for discrete nonlinear systems uses parallelotope bundles and Bernstein polynomials.
%
%Our main hypothesis in this paper is that generating templates in a dynamic manner would improve the accuracy of the reachable set.
%
To this end, we investigate two techniques for generating the template directions.
The first technique approximates the dynamics as a linear transformation and generates templates using this linear transformation.
The second technique uses Principal Component Analysis (PCA) of sample trajectories for generating templates.
We have implemented our approach in a Python-based tool called Kaa and improve its performance by two main enhancements.
%
% have implemented two main enhancements to it.
%
The tool is modular and use two types of global optimization solvers, the first using Bernstein polynomials and the second using NASA's Kodiak nonlinear optimization library.
Second, we leverage the natural parallelism of the reachability algorithm and parallelize the Kaa implementation.
% thus leveraging the cloud computing resources.
%
We demonstrate the improved accuracy of our approach on several standard nonlinear benchmark systems.
%
% In this paper, we describe Kaa, a terse Python implementation of reachable set computation which leverages the widely used symbolic package \emph{sympy}.
% %
% Additionally, we simplify the user interface and provide easy-to-use plotting utilities.
% %
% %
% We believe that our tool has pedagogical value given the simplicity of the implementation and its user-friendliness.
%

\end{abstract}

\section{Introduction}
\label{sec:intro}

One of the most widely-used techniques for performing safety analysis of nonlinear dynamical systems is reachable set computation.
The reachable set is defined to be the set of states visited by at least one of the trajectories of the system starting from an initial set.
Computing the reachable set for nonlinear systems is challenging primarily due to two reasons:
First, the tools for performing nonlinear analysis are not very scalable.
Second, computing the reachable set using set representations involves wrapping error.
That is, the overapproximation acquired at a given step would increase the conservativeness of the overapproximation for all future steps.

One of the techniques for computing the overapproximation of reachable sets for discrete time nonlinear systems is to use parallelotope bundles.
Here, the reachable set is represented as a parallelotope bundle, an intersection of several parallelotopes.
One of the advantages of this technique is its utilization of a special form of nonlinear optimization problem to overapproximate the reachable set.
The usage of a specific form of nonlinear optimization mitigates the drawback involved with the scalability of nonlinear analysis.

However, wrapping error still remains to be a problem for reachability using parallelotope bundles.
The template directions for specifying these parallelotopes are provided as an input by the user.
Often, these template directions are selected to be either the cardinal axis directions or some directions from octahedral domains.
However, it is not clear that the axis directions and octagonal directions are optimal for computing reachable sets.
Also, even an expert user of reachable set computation tools may not be able to provide a suitable set of template directions for computing reasonably accurate over-approximations of the reachable set.
Picking unsuitable template directions would only cause the wrapping error to increase, thus increasing the conservativeness of the safety analysis.

In this paper, we investigate techniques for generating template directions automatically and dynamically.
That is, instead of providing the template directions to compute the parallelotope, the user just specifies the number of templates and the algorithm automatically generates the template directions.
We study two techniques for generating the template directions.
First, we compute a local linear approximation of the nonlinear dynamics and use the linear approximation to compute the templates.
Second, we generate a specific set of sample trajectories from the set and use principal component analysis (PCA) over these trajectories.
We observe that the accuracy of the reachable set can be drastically improved by using templates generated using these two techniques.
For standard nonlinear benchmark systems, we show that generating templates in a dynamic fashion improves the accuracy of the reachable set by two orders of magnitude.
We demonstrate that even when the size of the initial set increases, our template generation technique returns more accurate reachable sets than both manually-specified and random template directions.

% [Introduction para]
% \begin{itemize}
% \item Reachable set computation.
% \item Nonlinear dynamics is challenging.
% \item Overapproximation, wrapping error.
% \item Parallelotope bundle reachability.
% \end{itemize}

% [Templates for reachability]
% \begin{itemize}
% \item Often the templates are generated statically.
% \item However, the most appropriate directions for templates that improve the accuracy is unknwon.
% \item Challenges: Wrapping error, cannot predict.
% \item Static template directions often aggrevate such errors.
% \item Generating templates dynamically is a challenge.
% \end{itemize}

% [Template generation]
% \begin{itemize}
% \item Template based overapproximation are used extensively in verification.
% \item More on this in the related work section.
% \item Dynamic templates using PCA have been investigated already.
% \item Not explicitly in the context of parallelotope bundles.
% \item In this paper we propose two techniques for  generating templates dynamically.
% \end{itemize}

\section{Related Work}

Reachable set computation of nonlinear systems using template polyhedra and Bernstein polynomials has been first proposed in~\cite{dang2009image}.
In~\cite{dang2009image}, Bernstein polynomial representation is used to compute an upper bound of a special type of nonlinear optimization problem~\cite{garloff2003bernstein}.
Several improvements to this algorithm were suggested in~\cite{dang2012reachability,sassi2012reachability} and~\cite{dang2014parameter} extends it for performing parameter synthesis.
The representation of parallelotope bundles for reachability was proposed in~\cite{dreossi2016parallelotope} and the effectiveness of using bundles for reachability was demonstrated in~\cite{dreossi2017sapo,dreossi2017reachability}.
However, all of these papers used static template directions for computing the reachable set.

Using template directions for reachable set has been proposed in~\cite{sankaranarayanan2008symbolic} and later improved in~\cite{dang2011template}.
Leveraging the principal component analysis of sample trajectories for computing reachable set has been proposed in~\cite{stursberg2003efficient,chen2011choice,seladji2017finding}.
More recently, connections between optimal template directions for reachability of linear dynamical systems and bilinear programming have been highlighted in~\cite{gronski2019template}.
For static template directions, octahedral domain directions~\cite{clariso2004octahedron} remain a popular choice.

\section{Preliminaries}
\label{sec:prelims}

The state of a system, denoted as $x$, lies in a domain $D \subseteq \reals^n$.
A discrete-time nonlinear system is denoted as
\begin{equation}
  x^{+} = f(x)
\label{eq:sys}
\end{equation}
where $f: \reals^{n} \rightarrow \reals^{n}$ is a nonlinear function.
The trajectory of a system that evolves according to Equation~\ref{eq:sys}, denoted as $\xi(x_0)$ is a sequence $x_0, x_1, \ldots $ where $x_{i+1} = f(x_i)$.
The $k^{th}$ element in this sequence $x_k$ is denoted as $\xi(x_0,k)$.
Given an initial set $\Theta \subseteq \reals^n$, the \emph{reachable set} at step $k$, denoted as $\Theta_k$ is defined as
\begin{equation}
  \Theta_k = \{ \xi(x,k)\: | \: x \in \Theta\}
\label{eq:reachset}
\end{equation}

A parallelotope $P$, denoted as a tuple $\tup{a, G}$ where $a \in \reals^n$ is called the \emph{anchor} and $G$ is a set of vectors $\{g_1, g_2, \ldots, g_n\}$, $\forall_{1 \leq i \leq n} ~ g_i \in \reals^n$ called \emph{generators}, represents the set
\begin{equation}
P = \{ x\:|\: \exists \alpha_1, \ldots, \alpha_n, \mbox{ such that }  0 \leq \alpha_i \leq 1, x = a + \sum_{i=1}^n \alpha_i g_i \}.
\label{eq:ptope}
\end{equation}

We call this representation as the \emph{generator representation} of the parallelotope.
We refer to a generator of a specific parallelotope $P$ using dot notation, for example $P.g_1$.
For readers familiar with zonotopes~\cite{girard2005reachability,althoff2010computing}, a parallelotope is a special form of zonotope where the number of generators $n$ equals the dimensionality of the set.
One can also represent the parallelotope as a conjunction of half-space constraints.
In half-space representation, a parallelotope is represented as a tuple $\tup{\T, c_{l}, c_{u}}$ where $\T \in \reals^{n \times n}$ are called \emph{template directions} and $c_{l}, c_{u} \in \reals^{n}$ such that $\forall_{1 \leq i \leq n} ~  c_{l}[i] \leq c_{u}[i]$ are called \emph{bounds}. The half-space representation defines the set of states
$$
P = \{\: x\: | \: c_{l} \leq \T x \leq c_{u} \}.
$$
%
% The rows in $\Lambda$ are called template directions in the parallelotope.
%
Intuitively, the $i^{th}$ constraint in the parallelotope corresponds to an upper and lower bound on the function $\T_{i} x$.
That is, $c_{l}[i] \leq \T_{i}x \leq c_{u}[i]$.
The half-plane representation of a parallelotope can be converted into the generator representation by computing $n+1$ vertices $v_1, v_2, \ldots, v_{n+1}$ of the parallelotope in the following way.
The vertex $v_1$ is obtained by solving the linear equation $\Lambda x = c_{l}$.
The $j+1$ vertex is obtained by solving the linear equation $\Lambda x = \mu_{j}$ where $\mu_{j}[i] = c_{l}[i]$ when $i \neq j$ and $\mu_{j}[j] = c_{u}[j]$.
The anchor $a$ of the parallelotope is the vertex $v_1$ and the generator $g_{i} = v_{i+1} - v_{1}$.

\begin{example}
  \label{ex:ptope}
  Consider the xy-plane and the parallelotope $P$ given in half-plane representation as $0 \leq x-y \leq 1$, $0 \leq y \leq 1$.
  This is a parallelotope with vertices at $(0,0)$, $(1,0)$, $(2,1)$, and $(1,1)$.
  In the half-space representation, the template directions of the parallelotope $P$ are given by the directions $[1, -1]$ and $[0, 1]$.
  The half-space representation in matrix form is given as follows:

  \begin{equation}
    \begin{bmatrix} 0 \\ 0 \end{bmatrix} \leq \begin{bmatrix}  1 & -1 \\ 0 &  1 \end{bmatrix}  \begin{bmatrix} x \\ y \end{bmatrix} \leq \begin{bmatrix} 1 \\ 1 \end{bmatrix}. \label{eq:ptopeexample}
\end{equation}

  To compute the generator representation of $P$, we need to compute the \emph{anchor} and the \emph{generators}.
  The anchor is obtained by solving the linear equations $x-y = 0, y = 0$.
  Therefore, the anchor $a$ is the vertex at origin $(0,0)$
  To compute the two generators of the parallelotope, we compute two vertices of the parallelotope.
  Vertex $v_1$ is obtained by solving the linear equations $x - y = 1, y = 0$.
  Therefore, vertex $v_1$ is the vertex $(1,0)$.
  Similarly, vertex $v_2$ is obtained by solving the linear equations $x-y = 0, y = 1$.
  Therefore, $v_2$ is the vertex $(1,1)$.
  The generator $g_1$ is the vector $v_1 - a$, that is $(1,0)- (0,0) = (1,0)$
  The generator $g_2$ is the vector $v_2 - a$, that is $(1,1) - (0,0) = (1,1)$.
  Therefore, all the points in the paralellotope can be written as $(x,y) = (0,0) + \alpha_1 (1,0) + \alpha_2(1,1)$, $0 \leq \alpha_1, \alpha_2 \leq 1$.
  %
  % We can also convert a generator representation of a parallelotope into the half-plane representation.
  %
\end{example}

A parallelotope bundle $Q$ is a set of parallelotopes $\{P_1, \ldots, P_m\}$.
The set of states represented by a parallelotope bundle is given as the intersection
\begin{equation}
Q = \bigcap_{i=1}^m P_i.
\end{equation}
Often, the various parallelotopes in a bundle share common template directions.
In such cases, the conjunction of all the parallelotope constraints in a bundle $Q$ is written as $c_{l}^Q \leq T^Q x \leq c_{u}^Q$.
Notice that the number of upper and lower bound half-space constraints in this bundle are stricly more than $n$ in these cases, i.e., $T^Q \in R^{m \times n}$ where $m>n$.
Each parallelotope in such a bundle is represented as a subset of constraints in $c_{l}^Q \leq T^Q x \leq c_{u}^Q$.
These types of bundles are often considered in the literature. \cite{dang2009image,dang2012reachability,dreossi2017reachability}

Alternatively, we consider parallelotope bundles where the consisting parallelotopes do not share template directions.
We consider such bundles because we generate the $n$ template directions automatically at each step.
%
% collectively generate one set of template directions.

The basic building block in this work is a conservative overapproximation to a constrained nonlinear optimization problem with a box domain.
Consider a nonlinear function $h : \reals^n \rightarrow \reals$ and the optimization problem denoted as $\mathsf{optBox(h)}$ as

\begin{eqnarray}
  \max ~ h(x) \label{eq:maxsup}\\
  s.t. ~~ x \in [0,1]^{n}.\nonumber
\end{eqnarray}

For computing the reachable set of a nonlinear system, we need an upper bound for the optimization problem.
Several techniques using interval arithmetic and Bernstein polynomials have been developed in the recent past \newline\cite{kodiak,garloff2003bernstein,munoz2013formalization,smith2009fast}.

\vspace{-1.2em}
\section{Reachability Algorithm}
\label{sec:theory}

In this work, we develop parallelotope reachability algorithms that are \textbf{automatic} with \textbf{dynamic} parallelotopes.
The state of the art, in contrast, is \textbf{manual}, where the user specifies a set of parallelotope directions at the beginning.
The parallelotopes are also \textbf{static} and do not change during the course of the computation.
In this section, we detail the modifications to the algorithm and present their correctness arguments.

\subsection{Manual Static Algorithm}
\label{sec:manualstatic}

We first present the original algorithm\cite{dang2012reachability} where the user manually specifies the number of parallelotopes and a set of static directions for each parallelotope.

Recall the system is $n$-dimensional with dynamics function $f: \reals^{n} \rightarrow \reals^{n}$.
The parallelotope bundle $Q$ is specified as a collection of $m$ template directions $\T^{Q} \in \reals^{m \times n} (m > n)$ and the set of constraints that define each of the member parallelotopes.
%

% consisting of $m$ parallelotopes , the user should manually provide $n$ generator directions for each parallelotope.
%
% We write the collective set of generator directions as $\mathcal{G}$, which is a set of sets, each of which contains $n$ direction vectors.
%
Another input to the algorithm is the initial set, given as a parallelotope $P_0$.
When the initial set is a box, $P_0$ consists has axis-aligned template directions.
The output of the algorithm is, for each step $k$, the set $\overline\Theta_k$, which is an overapproximation of the reachable set at step $k$, $\Theta_k \subseteq \overline\Theta_k$.

The high-level pseudo-code is written in Algorithm~\ref{alg:old}.
The algorithm simply calls \tbundle for each step, producing a new parallelotope bundle computed from the previous step's bundle.
To compute the image of $Q$, the algorithm computes the upper and lower bounds of $f(x)$ with respect to each template direction.
Since computing the maximum value of $f(x)$ along each template direction on $Q$ is computationally difficult, the algorithm instead computes the maximum value over each of the constituent parallelotopes and uses the minimum of all these maximum values.
The \tbundle operation works as follows.
Consider a parallelotope $P$ in the bundle $Q$.
From the definition, it follows that $Q \subseteq P$.
%
% and half-space representation as $\tup{\T, c_{l}, c_{u}$.
Given a template direction $\T_i$, the maximum value of $\T_{i} f(x)$ for all $x \in Q$ is less than or equal to the maximum value of $\T_{i} f(x)$ for all $x \in P$.
Similar argument holds for the minimum value of $T_{i} f(x)$ for all $x \in Q$.
%
% We would like to compute the image of the parallelotope $P$ with respect to the nonlinear transformation $f(x)$.
%
% That is, we would like to compute $P'$ such that $\forall x \in P, f(x) \in P'$.
%
% Given that current techniques use static template directions, the parallelotope $P'$ also uses the template directions $\T^{P}$.
%
To compute the upper and lower bounds of each template direction $\T_{i} f(x)$, for all $x \in P$, we perform the following optimization.
\begin{eqnarray}
  \max ~ \T_i^{P} \cdot f(x) \label{eq:maxf}\\
  s.t. ~~ x \in P.\nonumber
\end{eqnarray}

Given that $P$ is a parallelotope, all the states in $P$ can be expressed as a vector summation of anchor and scaled generators.
Let  $\tup{a, G}$ be the generator representation of $P$.
The optimization problem given in Equation~\ref{eq:maxf} would then transform as follows.

\begin{eqnarray}
  \max ~ \T_i \cdot f(a + \Sigma_{i=1}^{n} \alpha_i g_i) \label{eq:maxalpha}\\
  s.t. ~~ \overline\alpha \in [0,1]^{n}.\nonumber
\end{eqnarray}

Equation~\ref{eq:maxalpha} is a form of $\mathsf{optBox(\T_{i} \cdot f)}$ over $[0,1]^n$.
One can compute an upper-bound to the constrained nonlinear optimization by invoking one of the Bernstein polynomial or interval-arithmetic-based methods.
Similarly, we compute the lowerbound of $\T_{i}f(x)$ for all $x \in P$ by computing the upperbound of $-1 \times \T_{i}f(x)$.

We iterate this process (i.e., computing the upper and lower bound of $T_{i}f(x)$) for each parallelotope in the bundle $Q$.
%
% Observe that the role of the parallelotopes in the bundle is to compute an upper bound on the maximum value of $T_{i}f(x)$ by expanding the domain from $Q$ to the corresponding parallelotope $P$.
%
Therefore, the tightest upper bound on $T_{i}f(x)$ over $Q$ is the least of the upper bounds computed from each of the parallelotopes.
A similar argument holds for lower bounds of $T_{i}f(x)$ over $Q$.
Therefore, the image of the bundle $Q$ will be the bundle $Q'$ where the upper and lower bounds for templates directions are obtained by solving several constrained nonlinear optimization problems.

\begin{lemma}
\label{lem:correctness}
The parallelotope bundle $Q'$ computed using \tbundle (Algorithm~\ref{alg:old}) is a sound overapproximation of the image of bundle $Q$ w.r.t the dynamics $x^{+} = f(x)$.
\end{lemma}
% \begin{proof}
%   Suppose that parallelotope $P$ is part of the bundle $Q$ and is provided in generator representation as $\tup{a, G}$.
%   %
%   By definition, we know that $Q \subseteq P$.
%   % in generator representation and as $\tup{\T, c_{l}, c_{u}}$ in half-space representation.
%   %
%   Consider an $x \in Q$ and $x' = f(x)$.
%   %
%   Given a template direction $\T_{i}$, we know that $\T_{i} x' \leq \mathsf{optBox(\T_{i} \cdot f)}$ (Equation~\ref{eq:maxalpha}) and $-1\times optBox(-1\times \T_{i} \cdot f) \leq \T_{i} x'$.
%   %
%   Therefore, for all $x \in Q$, $c_{l}'[i] \leq \T_{i} \cdot f(x) \leq c_{u}'[i]$.
%   %
%   Therefore, for all $x \in Q$, $f(x) \in Q'$.
% \end{proof}

% \textcolor{red}{\textbf{SRIDHAR: add this (as well as argument for correctness)}}

% old algorithm static manual templates

\begin{algorithm}[t]
  \SetKwInput{KwInput}{Input}
  \SetKwInput{KwOutput}{Output}
  \SetKwFunction{AppendPCA}{AppendNewPCATemplates}
  \SetKwFunction{AppendLin}{AppendNewLinAppTemplates}
  \SetKwFunction{TransformBund}{TransformBundle}
  \SetKwFunction{UpdateTemp}{UpdateTemplates}
  \SetKwFunction{GetSupp}{GetSupportPoints}
  \SetKwFunction{PropPoints}{PropagatePointsOneStep}
  \SetKwFunction{PCA}{PCA}
  \SetKwFunction{ApproxLinearTrans}{ApproxLinearTrans}
  \SetKwFunction{SetLifeSpan}{SetLifeSpan}
  \SetKwFunction{AddTemptoBund}{AddTemplateToBundle}
  \SetKwFunction{RemoveTemp}{RemoveTempFromBund}
  \SetKwProg{Fun}{Proc}{:}{}

\SetAlgoLined
\DontPrintSemicolon

\KwInput{Dynamics $f$, Initial Parallelotope $P_0$, Step Bound $S$, Template Dirs $\T$, indexes for parallelotopes $I$}
\KwOutput{Reachable Set Overapproximation $\overline\Theta_k$ for each step $k$}
$Q_0 = \{ P_0 \}$ \;
 \For{$k \in [1, 2, \ldots, S]$}{
    $Q_k$ = \TransformBund($f$, $Q_{k-1}$, $\T$) \;
   $\overline\Theta_k = Q_k$ \;
 }
 \Return{$\overline\Theta_1 \ldots \overline\Theta_S$} \;
 \;
 \Fun{\TransformBund{$f$, $Q$, $\T$}}{
   $Q' \gets \{\}$; $c_{u} \gets +\infty$; $c_{l} \gets -\infty$\;
   \For{each parallelotope $P$ in $Q$}{
     % $\tup{\T, c_{l}, c_{u}} \gets \mathsf{half-spaceRepresentation}(P)$\;
     $\tup{a, G} \gets \mathsf{generatorRepresentation}(P)$\;
     \For{each template direction $\T_i$ in the template directions $\T$ }{
       $c_{u}'[i] \gets \mathsf{min}\{ \mathsf{optBox(\T_{i} \cdot f)}, c_{u}'[i] \}$ (Equation~\ref{eq:maxalpha})\;
       $c_{l}'[i] \gets \mathsf{max}\{ -1 \times \mathsf{optBox(-1\times \T_{i} \cdot f)}, c_{l}'[i] \}$\;
     }
   }
   Construct parallelotopes $P_{1}', \ldots, P_{k}'$ from $\T, c_{l}', c_{u}'$ and indexes from $I$\;
   % \textcolor{red}{\textbf{SRIDHAR: add this}} \;
   $Q' \gets \{P_{1}', \ldots, P_{k}'\}$\;
     % $P' \gets \tup{\T, c_{l}', c_{u}'}$\;
     % $Q' \gets Q' \cup \{P'\}$\;
   \Return{$Q'$} \;
 }
 \caption{Reachable set computation using manual and static templates.
   % Manual, Static Reachability Algorithm
 }
\label{alg:old}
\end{algorithm}

\vspace{-1.5em}
\subsection{Automatic Dynamic Algorithm}

The proposed automatic dynamic algorithm does not require the user to provide the set of template directions $\T$; instead it generates these templates directions automatically at each step.
We use two techniques to generate such template directions, first: computing local linear approximations of the dynamics and second, performing principal component analysis (PCA) over sample trajectories.
To do this, we first sample a set of points in the parallelotope bundle called \emph{support points} and propagate them to the next step using the dynamics $f$.
Support points are a subset of the vertices of the parallelotope that either maximize or minimize the template directions.

Intuitively, linear approximations can provide good approximations when the dynamics function is a time-discretization of a continuous system.
In this case, for small time steps a nonlinear function can be approximated fairly accurately by a linear transformation.
We use the support points as a data-driven approach to find the best-fit linear function to use.
If the dynamics of a system is linear, i.e., $x^{+} = Ax$, the image of the parallelotope $c_{l} \leq \T x \leq c_{u}$, is the set $c_{l} \leq \T \cdot A^{-1} x \leq c_{u}$.
Therefore, given the template directions of the initial set as $\T_0$, we compute the local linear approximation of the nonlinear dynamics and change the template directions by multiplying them with the inverse of the approximate linear dynamics.
The second technique for generating template directions performs principal component analysis over the images of the support points.
Using PCA is a reasonable choice as it produces orthonormal directions that can construct a rotated box for bounding the points.

Observe that in general, the dynamics is nonlinear and therefore, the reachable set could be non-convex.
On the other hand, a parallelotope bundle is always a convex set.
To mitigate this discrepancy, we can improve accuracy of this representation by considering more template directions.
For this purpose, we use a notion of \emph{template lifespan}, where we use the linear approximation and/or PCA template directions not only from the current step, but also from the previous $L$ steps.
We will demonstrate the effectiveness and tune each of the options (PCA / linear approximation as well as lifespan option) in our evaluation.

The new approach is given in Algorithm~\ref{alg:new}.
In this algorithm, instead of fixing the set of templates, we compute one set of templates (that is, a collection of $n$ template directions), using linear approximation of the dynamics and PCA.
The algorithm makes use of helper function \texttt{hstack}, which converts column vectors into a matrix (as shown in Equation~\ref{eq:ptopeexample} provided in Example~\ref{ex:ptope}).
The notation $M_{*,i}$ is used to refer to the $i^{th}$ column of matrix $M$.
The \texttt{Maximize} function takes in a parallelotope bundle $Q$ and direction vector $v$ (one of the template directions), and returns the point $p \in Q$ that maximizes the dot product $v \cdot p$ (for computing support points).
This can be computed efficiently using linear programming.
The \texttt{ApproxLinearTrans} function computes the best approximation of a linear transformation given a list of points before and after the one-step transformation $f$.
More specifically, given a matrix $X$ of points before applying the transformation $f$, a matrix of points after the transformation $X'$, we perform a least-squares fit for the linear transition matrix $A$ such that $X' \approx AX$.
This can be computed by $A = X' X^\dagger$, where $X^\dagger$ is the Moore-Penrose pseudoinverse of $X$.
The \texttt{PCA} function returns a set of orthogonal directions using principal component analysis of a set of points.
Finally, \texttt{TransformBundle} is the same as in Algorithm~\ref{alg:old}.

\begin{algorithm}[t]
  \SetKwInput{KwInput}{Input}
  \SetKwInput{KwOutput}{Output}
  \SetKwFunction{CreatePCA}{CreatePCA}
  \SetKwFunction{CreateLin}{CreateLin}
  \SetKwFunction{TransformBund}{TransformBundle}
  \SetKwFunction{UpdateTemp}{UpdateTemplates}
  \SetKwFunction{GetSupp}{GetSupportPoints}
  \SetKwFunction{PropPoints}{PropagatePointsOneStep}
  \SetKwFunction{PCA}{PCA}
  \SetKwFunction{ApproxLinearTrans}{ApproxLinearTrans}
  \SetKwFunction{SetLifeSpan}{SetLifeSpan}
  \SetKwFunction{ExtractDirections}{ExtractDirections}
  \SetKwFunction{AddTemptoBund}{AddTemplateToBundle}
  \SetKwFunction{Maximize}{Maximize}
  \SetKwFunction{hstack}{hstack}
  \SetKwFunction{RemoveTemp}{RemoveTempFromBund}
  \SetKwProg{Fun}{Proc}{:}{}

\SetAlgoLined
\DontPrintSemicolon

\KwInput{Dynamics $f$, Initial Parallelotope $P_0$, Step Bound $S$}
\KwOutput{Reachable Set Overapproximation $\overline\Theta_k$ at each step $k$}

$Q_0 = \{P_0 \}$ \;

% $\T_0 = \{ \{ P_0.\T_1, \ldots P_0.\T_n \} \}$ \;
$\T = \hstack(P_0.\T_1, \ldots, P_0.\T_n)$ \tcp{Init Template Directions}

 \For{$k \in [1, 2, \ldots, S]$}{
    $P_{supp}$ = \GetSupp($Q_{k-1}$) (support points of $Q_{k-1}$) \;
    $P_{prop}$ = \PropPoints($P_{supp}$, $f$) (image of support points) \;

    $A$ =  \ApproxLinearTrans($P_{supp}$, $P_{prop}$) \; \label{ln:linearapprox}
    $\T = \T \cdot A^{-1}$ \;

    $\T_k^\text{lin} = \{ \{\T_{*,1} , \ldots, \T_{*,n} \} \} $ \; \label{ln:linapp}
    $\T_k^\text{pca} = \{ \PCA(P_{prop}) \} $\; \label{ln:pca}

    $\T_k = \T_k^\text{lin} \cup \T_k^\text{pca}$ \;
    \;
    \tcc{For lifespan $L$, instead call \TransformBund with $\T_k \cup \T_{k-1} \cup \ldots \cup \T_{k-L}$}
    $Q_k$ = \TransformBund($f$, $Q_{k-1}, \T_k$) \;
   $\overline\Theta_k \gets Q_k$ \;
 }
 \Return{$\overline\Theta_1 \ldots \overline\Theta_S$} \;
 \;
   \Fun{\GetSupp{$Q$}}{
     $P_{supp} = \emptyset$ \;
      \For{$P \in Q$}{
        \For{$i \in [1, 2, \ldots, n]$}{
         $P_{supp} = P_{supp} \cup ~ \Maximize (Q, P.T_i) \cup ~ \Maximize (Q, -P.T_i)$
        }
     }

  \Return{$P_{supp}$}
  }

 \caption{Automatic, Dynamic Reachability Algorithm}
  \label{alg:new}
\end{algorithm}

Algorithm~\ref{alg:new} computes the dynamic templates for each time step $k$.
Line~\ref{ln:linearapprox} computes the linear approximation of the nonlinear dynamics and this linear approximation is used to compute the new template directions according to this linear transformation in Line~\ref{ln:linapp}.
The PCA directions of the images of support points is computed in line~\ref{ln:pca}.
For the time step $k$, the linear and PCA templates are given as $\T_{k}^{lin}$ and $\T_{k}^{pca}$, respectively.
To improve the accuracy of the reachable set, we compute the overapproximation of the reachable set with respect to not just the template directions at the current step, but with respect to other template directions for time steps that are within the lifespan $L$.
%
%This is because, the nonlinear systems we deal with are often %not chaotic in nature.
%
%Therefore, the templates at time step $k-1$ are very similar %to that of template directions at time step $k$.
%

\vspace{-1em}
\section{Evaluation}
\label{sec:eval}
%\vspace{-0.5em}
 We evaluate the efficacy of our dynamic parallelotope bundle strategies with our tool, \emph{Kaa} \cite{kim2020kaa}. Kaa is written in Python and relies on the \emph{numpy} library for matrix computations, \emph{sympy} library for all symbolic subsitution, and \emph{scipy}, \emph{matplotlib} for plotting the reachable sets and computing the volume for lower-dimensional systems. The optimization procedure for finding the direction offets is performed through the \emph{Kodiak} library. Finally, parallelization of the offset calculation procedures is implemented through the \emph{multiprocessing} module. To estimate volume of reachable sets, we employ two techniques for estimating volume of individual parallelotope bundles. For systems of dimension fewer than or equal to three, we utilize scipy's convex hull routine.
For higher-dimensional systems, we employ the volume of the tightest enveloping box around the parallelotope bundle.
The total volume estimate of the overapproximation will be the sum of all the bundles' volume estimates.

 %For two-dimensional models, note that there are only five non-trivial templates. The trivial template here refers to the initial box which is defined exactly the axis-aligned directions. Refer to the non-axis aligned templates. At each every time, pca. lin app directions to
\vspace{-1em}
\subsubsection{Model Dynamics}
For benchmarking, we select six non-linear models with polynomial dynamics. Benchmarks against more general dynamics can be found in the appendix of the \href{https://arxiv.org/abs/2105.11796}{expanded verison}. Many of these models are also implemented in \emph{Sapo} \cite{dreossi2017sapo}, a previous tool exploring reachability with {\bf static} parallelotope bundles. In these cases, we directly compare the performance of our dynamic strategies with the Sapo's static parallelotopes. To provide meaningful comparisions, we set the number of dynamic parallelotopes to be equal to the number of static ones excluding the initial box. Here, {\bf diagonal directions} are defined to be vectors created by adding and subtracting distinct pairs of unit axis-aligned vectors from each other. By {\bf diagonal parallelotopes}, we refer to parallelotopes defined only by axis-aligned and diagonal directions. Similarly, {\bf diagonal parallelotope bundles} are parallelotope bundles solely consisting of diagonal parallelotopes. Sapo primarily utilizes {\bf static diagonal parallelotope bundles} to perform its reachability computation.
Note that the initial box, which is defined only through the axis-aligned directions, is contained in every bundle.
For our experiments, we are concerned with the effects of additional static or dynamic parallelotopes added alongside the initial box. We refer to these parallelotopes as {\bf non-axis-aligned parallelotopes}.

%Maybe mention something about role of diagonal templates in program verfication here.
\begin{example}
In two dimensions, $\mathbb{R}^2$, we have the two unit axis-aligned directions, $[1,0]^T, [0,1]^T$. The diagonal directions will then be
\[ [1,1]^T, \; [1,-1]^T\]
Consequently, the diagonal parallelotopes will precisely be defined by unique pairs of these directions, giving us a total ${4 \choose 2} = 6$ diagonal parallelotopes.
% Similarly, in $\mathbb{R}^3$, we have three unit axis-aligned directions $[1,0,0]^T, [0,1,0]^T,[0,0,1]^T$. Calculating the diagonal directions yields the following directions:
% \[ [1,1,0]^T, \; [1,-1,0]^T, \; [1,0,1]^T,\;  [1,0,-1]^T,\;  [0,1,1]^T, \; [0,1,-1]^T \]
% Thus, there are ${9 \choose 3} = 84$ diagonal parallelotopes.
%
% Note that for both examples, we do not consider the directions' negative counterparts since we define parallelotopes use both the positive and negative counterparts of a its chosen directions.
%
% For two dimensions, the non-axis-aligned parallelotopes would be all parallelotopes excluding the one defined by directions $[0,1]^T, [1,0]^T$. In particular, there are 5 non-axis-aligned diagonal parallelotopes.
\end{example}
Table \ref{tab:modeldyns} summarizes five standard benchmarks used for experimentation. The last seven-dimensional COVID supermodel is explained in the subsequent subsection below.

\vspace{-1em}
\begin{table}[h!]
%\hspace{-5em}
  \centering
\begin{tabular}{|p{1.5cm}|c|p{1.7cm}|c|c|p{5cm}|}
\hline
Model & Dimension & Parameters & \# steps & $\Delta$ & \hspace{1.5cm}Initial Box \\
\hline
Vanderpol & 2 & \quad \quad \; - & 70 steps & 0.08 & $x \in [0,0.1], y \in [1.99,2]$ \\
\hline
Jet Engine& 2 & \quad \quad \; - & 100 steps & 0.2 & $x \in [0.8,1.2], y \in [0,8,1.2]$ \\
\hline
Neuron \cite{fitzhugh1961impulses}& 2 & \quad \quad \; - & 200 steps & 0.2 & $x \in [0.9,1.1], y \in [2.4,2.6]$ \\
\hline
SIR& 3 & $\beta=0.05$ \newline $\gamma=0.34$ & 150 steps & 0.1 & $s \in [0.79,0.8], i \in [0.19,0.2], r = 0$ \\
\hline
Coupled \newline Vanderpol & 4 & \quad \quad \; - & 40 steps & 0.08 & $x1 \in [1.25, 2.25], y1 \in [1.25, 2.25]$ \newline $x2 \in [1.25, 2.25], y2 \in [1.25, 2.25]$ \\
\hline
COVID & 7 & $\beta=0.05$ \newline $\gamma=0.0$ \newline $\eta=0.02$ & 200 steps & 0.08 & \quad \quad \quad \; \; Stated Below\\
\hline
\end{tabular}
\caption{Benchmark models and relevant information}
\label{tab:modeldyns}
\end{table}

\vspace{-2em}
\noindent \textbf{COVID Supermodel:}
We benchmark our dynamic strategies with the recently introduced COVID supermodel \cite{ansumali2020modelling}, \cite{indiansuper2020supermodel}. This model is a modified SIR model accounting for the possibility of \emph{asymptomatic} patients. These patients can infect susceptible members with a fixed probability. The dynamics account for this new group and its interactions with the traditional SIR groups.

\begin{align}
  \begin{split}
   S_A' & = S_A  -(\beta S_A(A+I))\cdot \Delta \\
   S_I' & = S_I  -(\beta S_I (A + I))\cdot \Delta \\
   A' & = A + (\beta S_I(A+I) - \gamma I)\cdot \Delta \\
   I' & = I + (\beta S_I (A+I) - \gamma I)\cdot  \Delta \\
   R_A' & = R_A + (\gamma A)\cdot \Delta \\
   R_I' & = R_I + (\gamma I)\cdot \Delta \\
   D' & = D + (\eta I)\cdot \Delta
  \end{split}
\end{align}
where the variables denote the fraction of a population of individuals designated as \emph{Susceptible to Asymptomatic $(S_A)$}, \emph{Susceptible to Symptomatic $(S_I)$}, \emph{Asymptomatic (A)}, \emph{Symptomatic (I)}, \emph{Removed from Asymptomatic $(R_A)$}, \emph{Removed from Symptomatic $(R_I)$}, and \emph{Deceased (D)}. We choose the parameters ($\beta = 0.25, \gamma=0.02, \eta=0.02$) where $\beta$ is the probablity of infection, $\gamma$ is the removal rate, and $\eta$ is the mortality rate. The parameters are set based on figures shown in \cite{ansumali2020modelling}. The discretization step is chosen to be $\Delta = 0.1$ and the initial box is set to be following dimensions: $S_A  \in [0.69, 0.7], \, S_I \in [0.09, 0.1], \, A \in [0.14, 0.15], \, I \in [0.04, 0.05], \, R_A  = 0,\, R_I  = 0, \, D  = 0$.

\vspace{-1em}
\subsubsection{Accuracy of Dynamic Strategies}
The results of testing our dynamic strategies against static ones are summarized in Table~\ref{tab:voltable}. For models previously defined in Sapo, we set the static parallelotopes to be exactly those found in Sapo.
If a model is not implemented in Sapo, we simply use the static parallelotopes defined in a model of equal dimension. To address the unavailability of a four-dimensional model implemented in Sapo, we sampled random subsets of five static non-axis-aligned parallelotopes and chose the flowpipe with smallest volume.
%
% As justifiction for this scheme, we wish to remark the time required in testing all possible subsets of diagonal parallelotopes.
%
A cursory analysis shows that the number of possible templates with diagonal directions grows with $O(n^n)$ with the number of dimensions and hence an exhaustive search on optimal template directions is infeasible.

% are $2{n \choose 2}$ diagonal directions for an $n$-dimensional system, giving a total of $n + 2 {n \choose 2}$ axis-aligned and diagonal directions. Thus, there are ${n + 2 {n \choose 2} \choose n} = {n^2 \choose n}$ possible diagonal parallelotopes for an $n$-dimensional system. This explosive growth in search space size compelled us to proceed by sampling the search space instead.

\begin{figure}[h!]
\begin{subfigure}{0.5\textwidth}
\includegraphics[width=\textwidth]{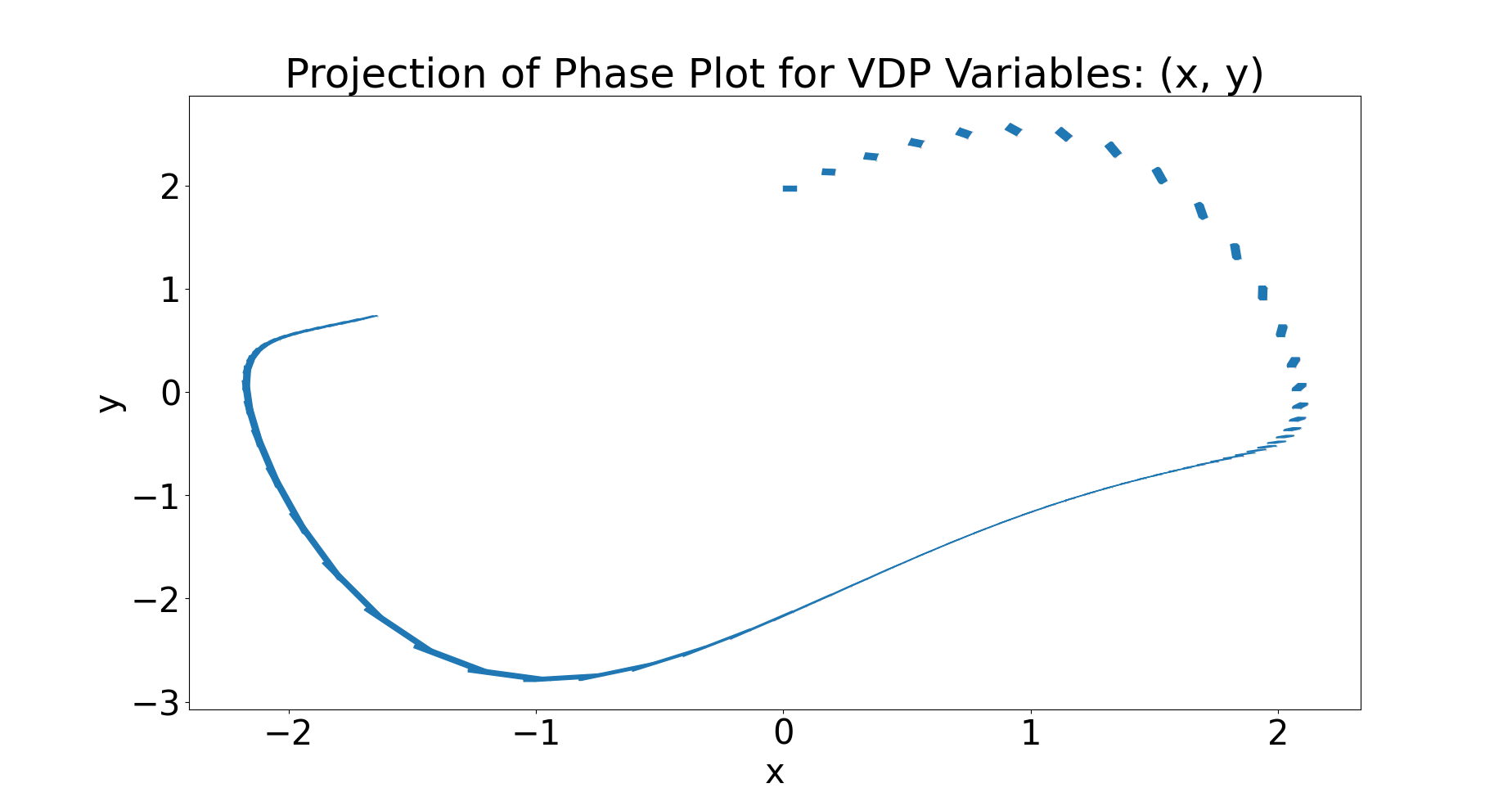}
\caption{5 Lin}
\end{subfigure}%
\begin{subfigure}{0.5\textwidth}
\includegraphics[width=\textwidth]{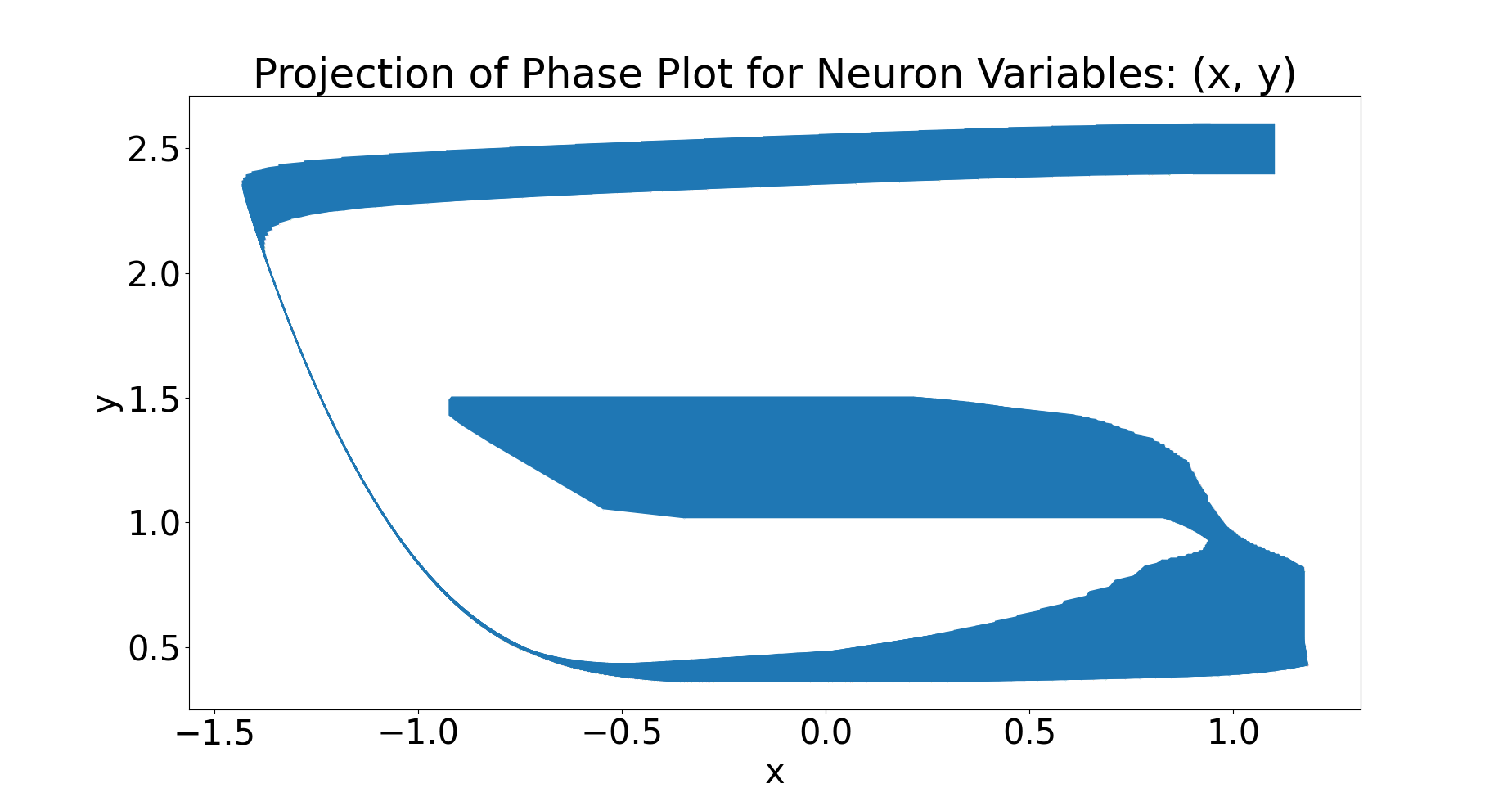}
\caption{1 PCA 5 Lin}
\end{subfigure}%

\begin{subfigure}{0.5\textwidth}
\includegraphics[width=\textwidth]{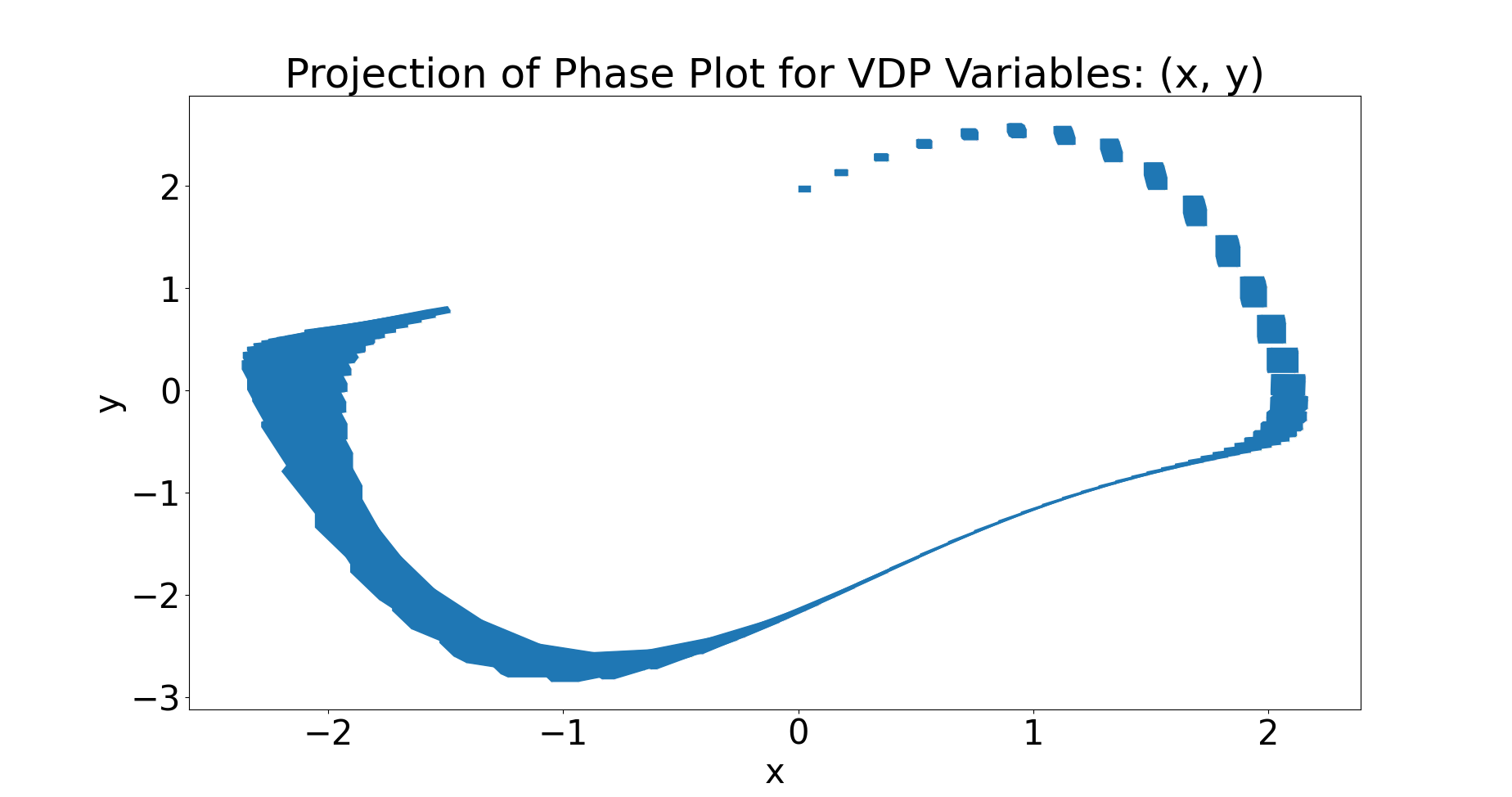}
\caption{5 PCA}
\end{subfigure}%
\begin{subfigure}{0.5\textwidth}
\includegraphics[width=\textwidth]{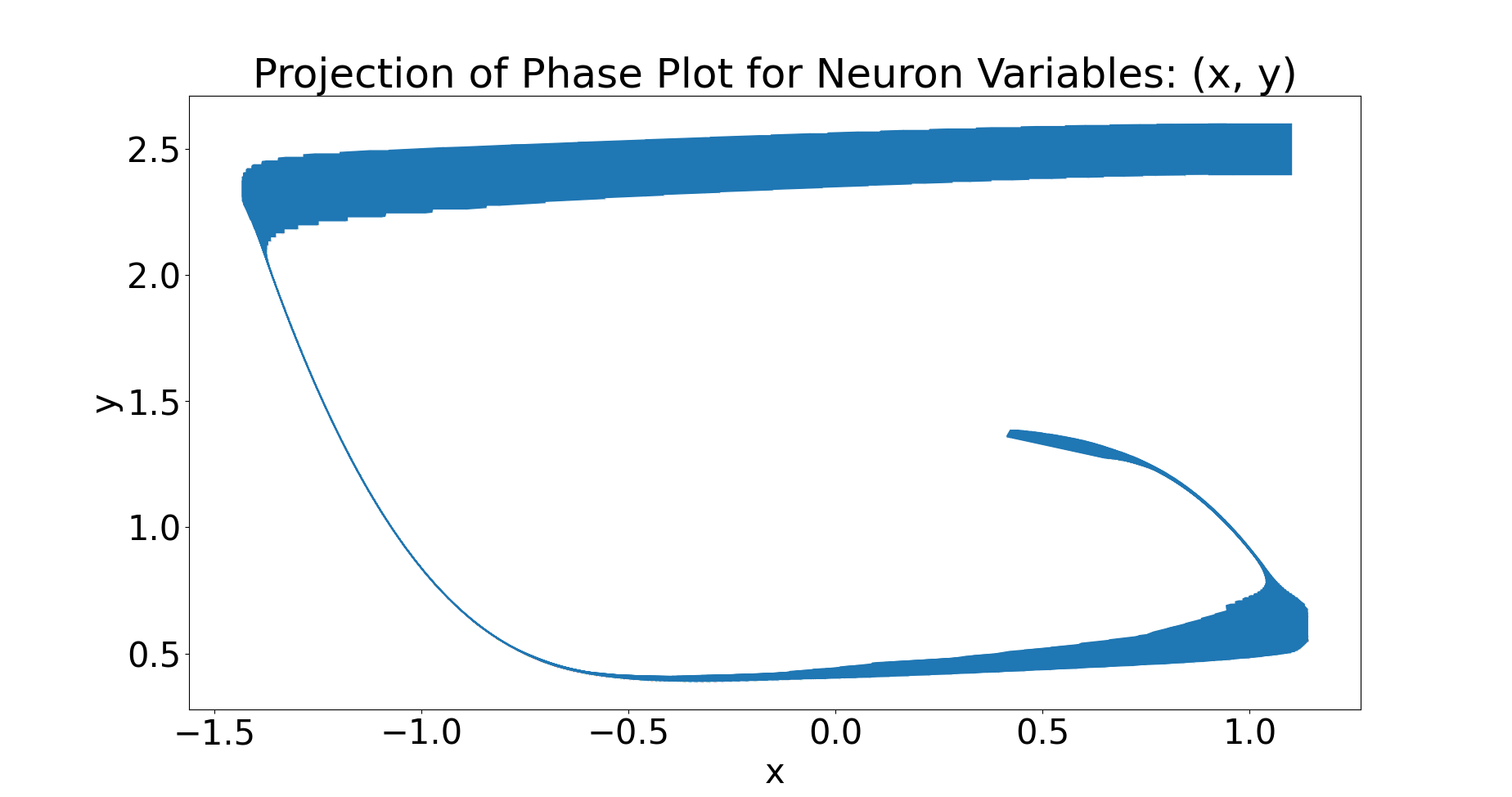}
\caption{5 PCA 1 Lin}
\end{subfigure}

\begin{subfigure}{0.5\textwidth}
\includegraphics[width=\textwidth]{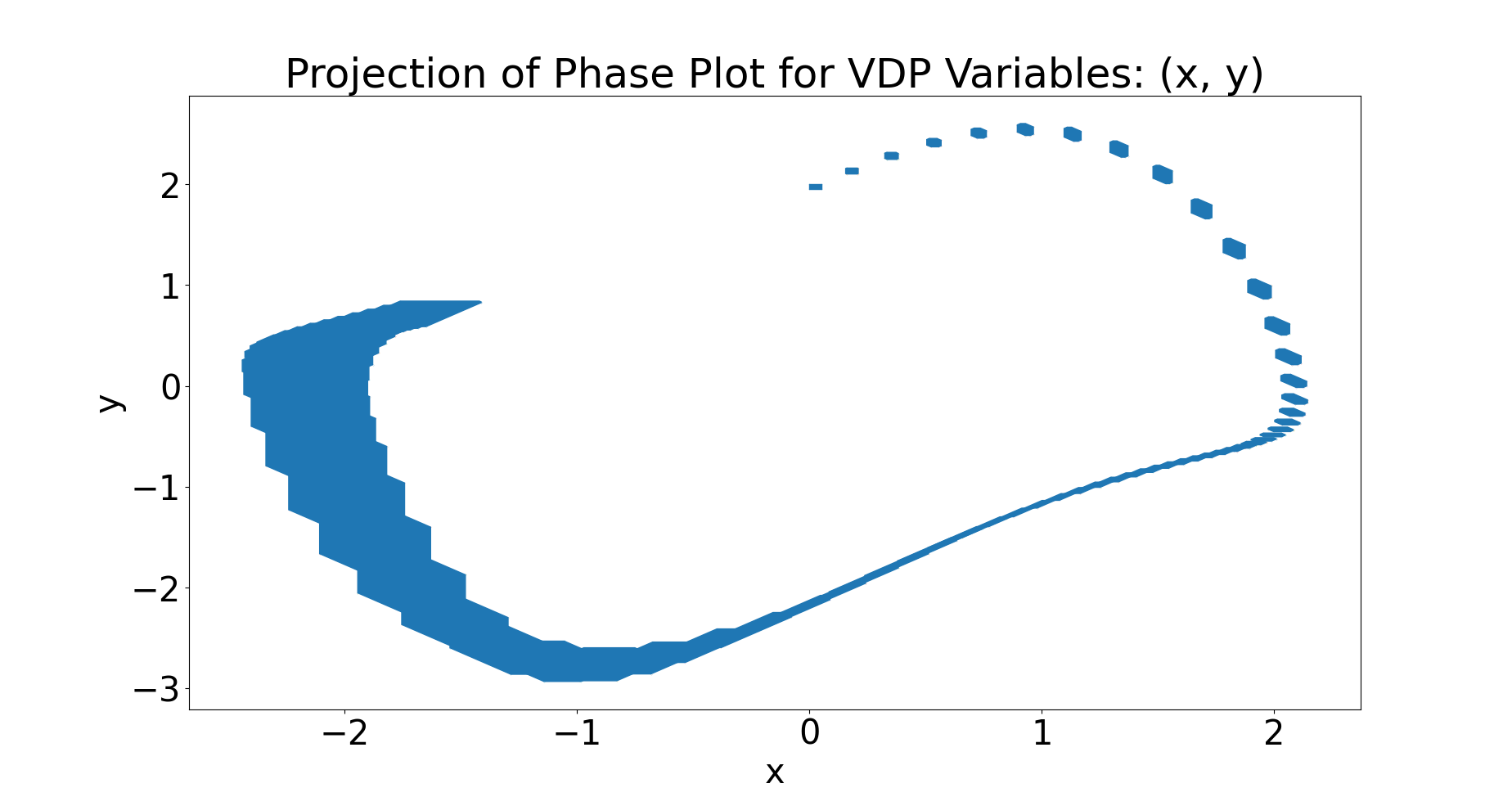}
\caption{Sapo}
\end{subfigure}%
\begin{subfigure}{0.5\textwidth}
\includegraphics[width=\textwidth]{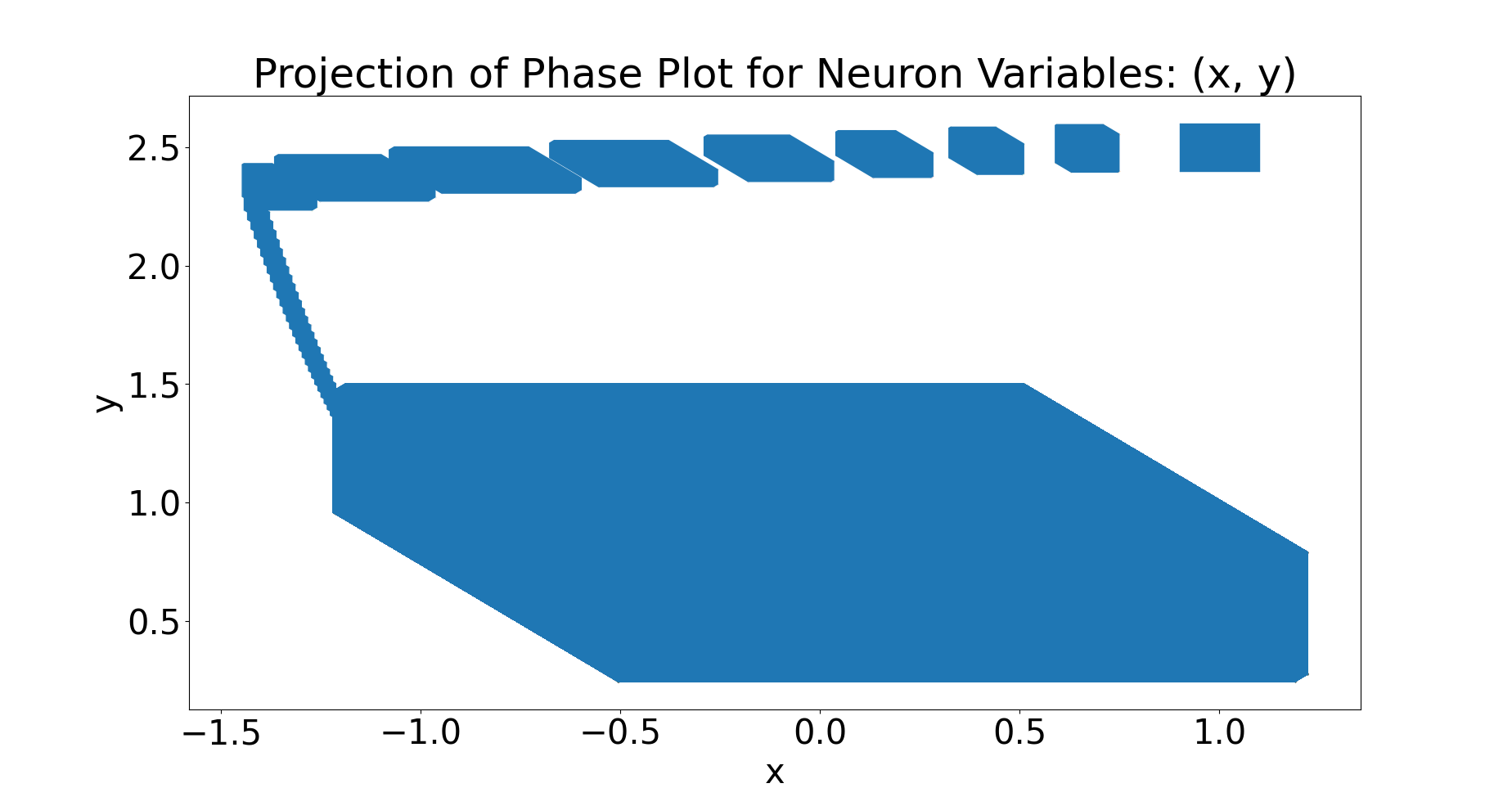}
\caption{Sapo}
\end{subfigure}
\caption{Effect of varying ratio between the number of PCA and Linear Approximation parallelotopes. The Vanderpol (left) and the FitzHugh-Nagumo Neuron (right) phase plots are shown to illustrate differing effects of varying the PCA/LinApp ratio. The initial set for the Vanderpol model is set to $x \in [0,0.05], \, y \in [1.95,2]$}
\label{fig:PCALinAppRatio}
\end{figure}

From our experiments, we conclude there is no universal optimal ratio between the number of dynamic parallelotopes defined by PCA and Linear Approxiation directions which perform well on all benchmarks. In Figure \ref{fig:PCALinAppRatio}, we demonstrate two cases where varying the ratio imparts differing effects. Observe that using parallelotopes defined by linear approximation directions is more effective than those defined by PCA directions in the Vanderpol model whereas the Neuron model shows the opposite trend.

\vspace{-1em}
\subsubsection{Performance under Increasing Initial Sets}
\begin{figure}[h!]
    \hspace{-1.5em}
    \begin{subfigure}{0.5\textwidth}
    \centering
    \includegraphics[width=1.1\textwidth, height=0.75\textwidth]{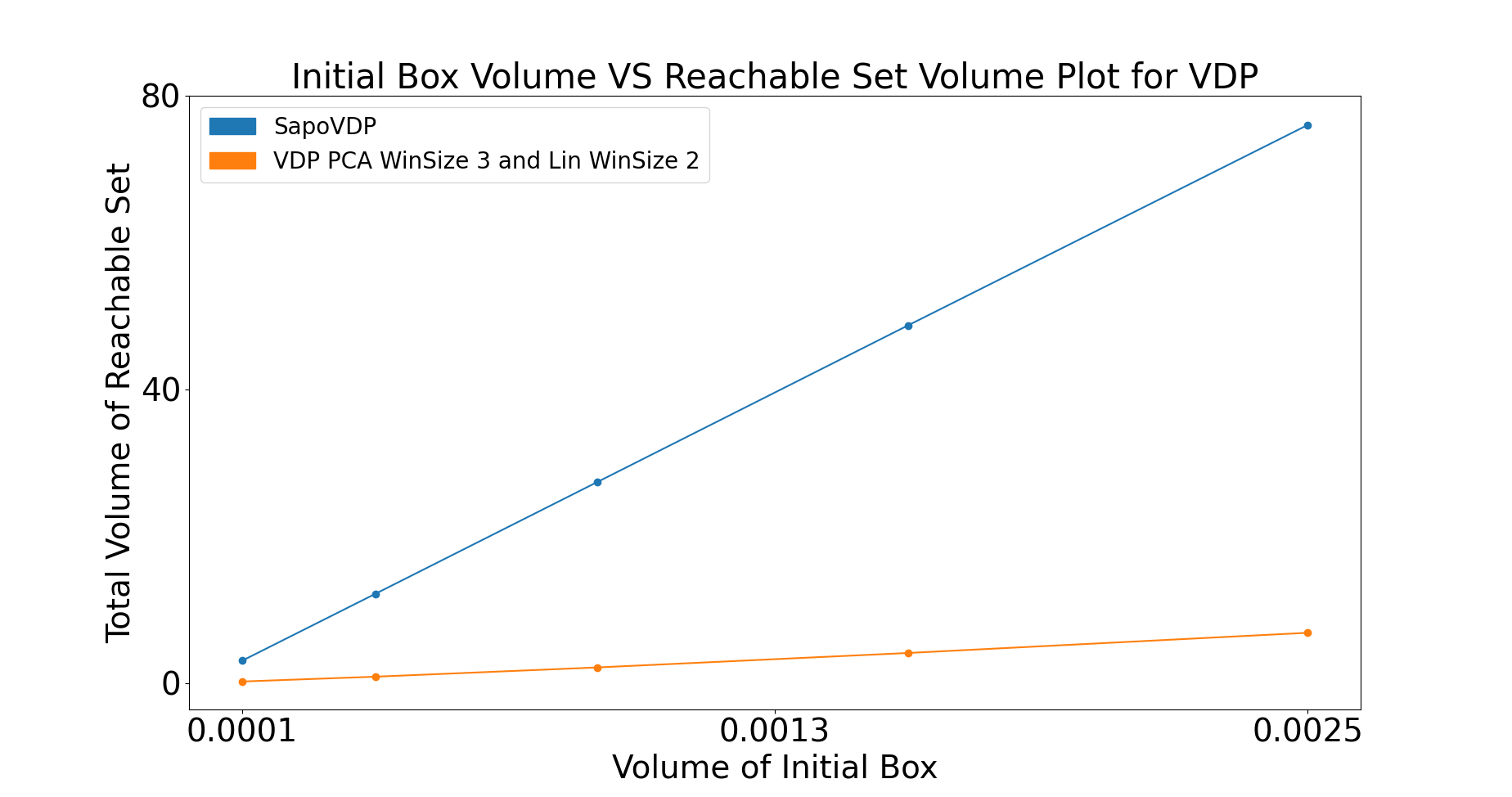}
    \caption{Vanderpol}
    \end{subfigure}%
    %\hspace{1em}
    \begin{subfigure}{0.5\textwidth}
    \centering
    \includegraphics[width=1.1\textwidth, height=0.75\textwidth]{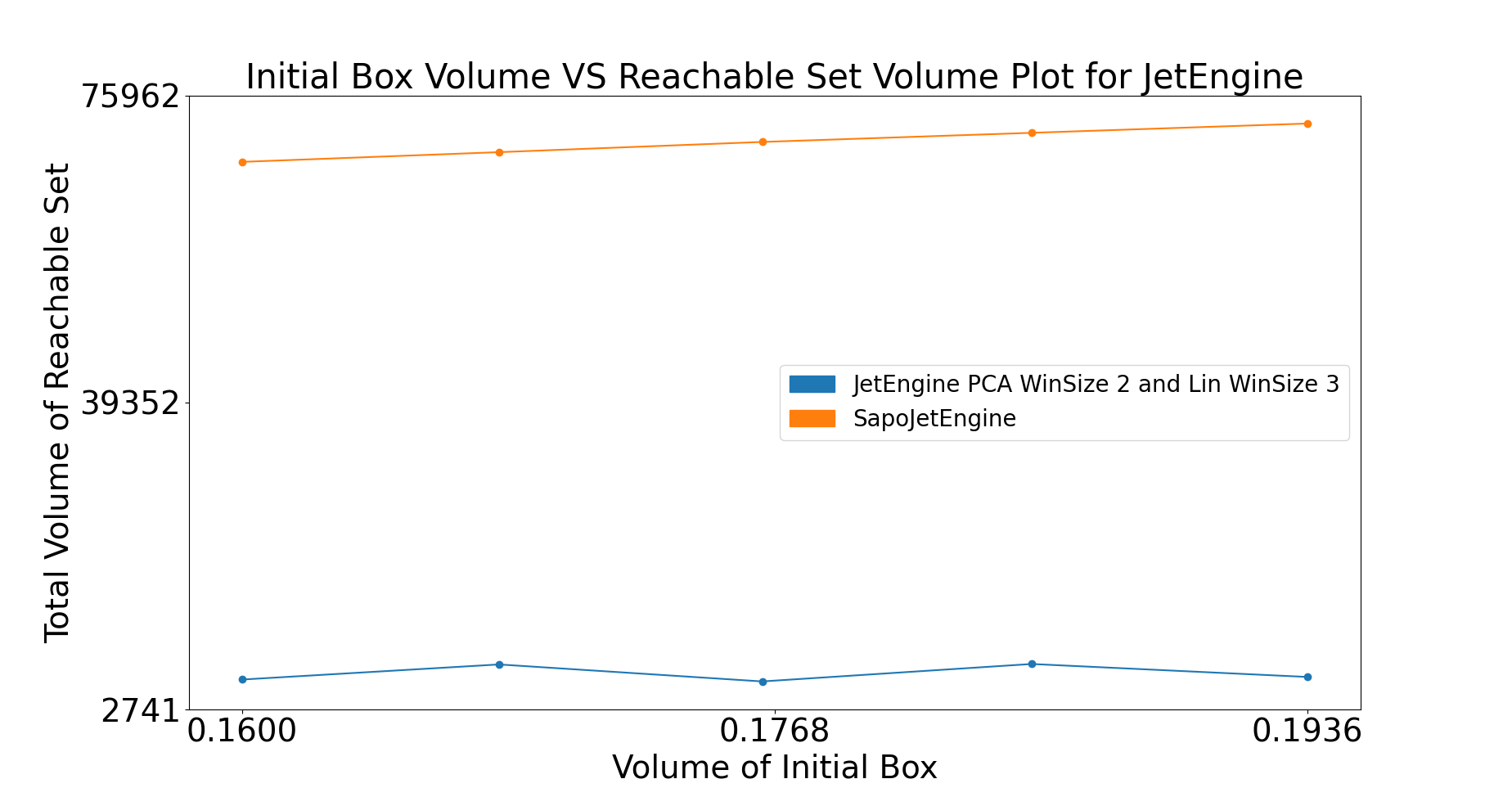}
    \caption{Jet Engine}
    \end{subfigure}

    \hspace{-1.5em}
    \begin{subfigure}{0.5\textwidth}
    \centering
    \includegraphics[width=1.1\textwidth, height=0.75\textwidth]{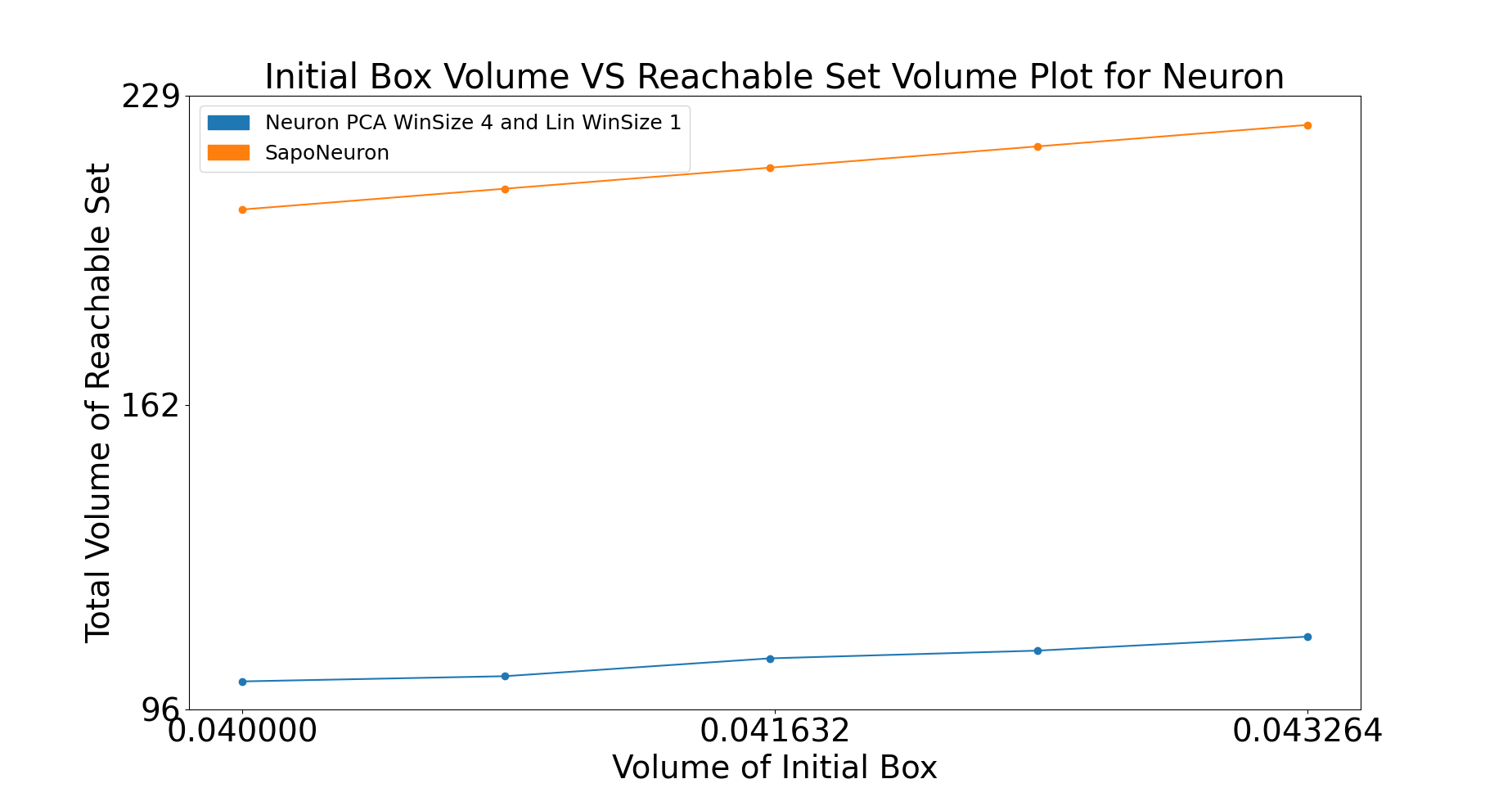}
    \caption{Neuron}
    \end{subfigure}%
    %\hspace{1em}
    \begin{subfigure}{0.5\textwidth}
    \centering
    \includegraphics[width=1.1\textwidth, height=0.75\textwidth]{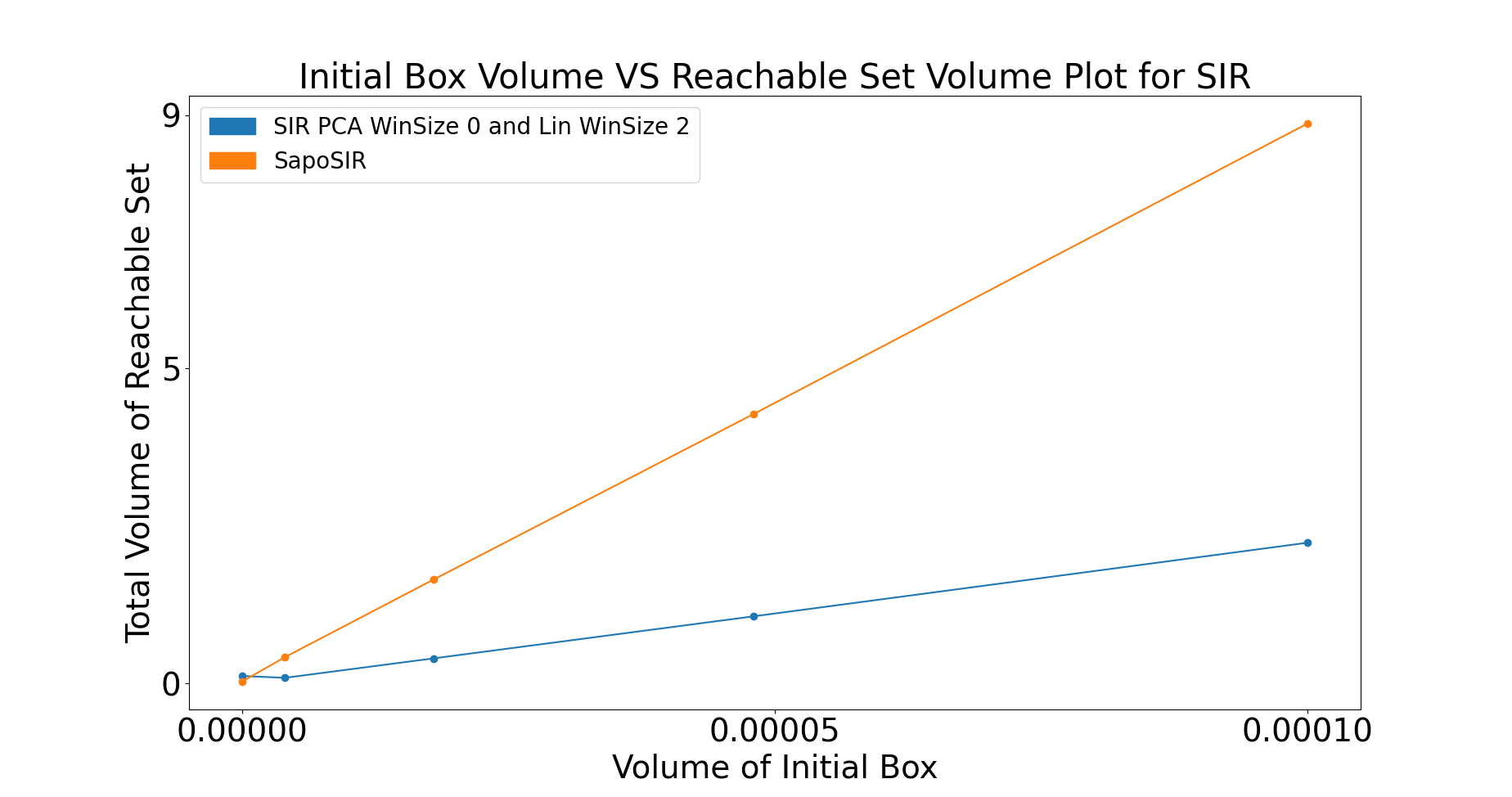}
    \caption{SIR}
    \end{subfigure}

    \hspace{-1.5em}
    \begin{subfigure}{0.5\textwidth}
    \centering
    \includegraphics[width=1.1\textwidth, height=0.75\textwidth]{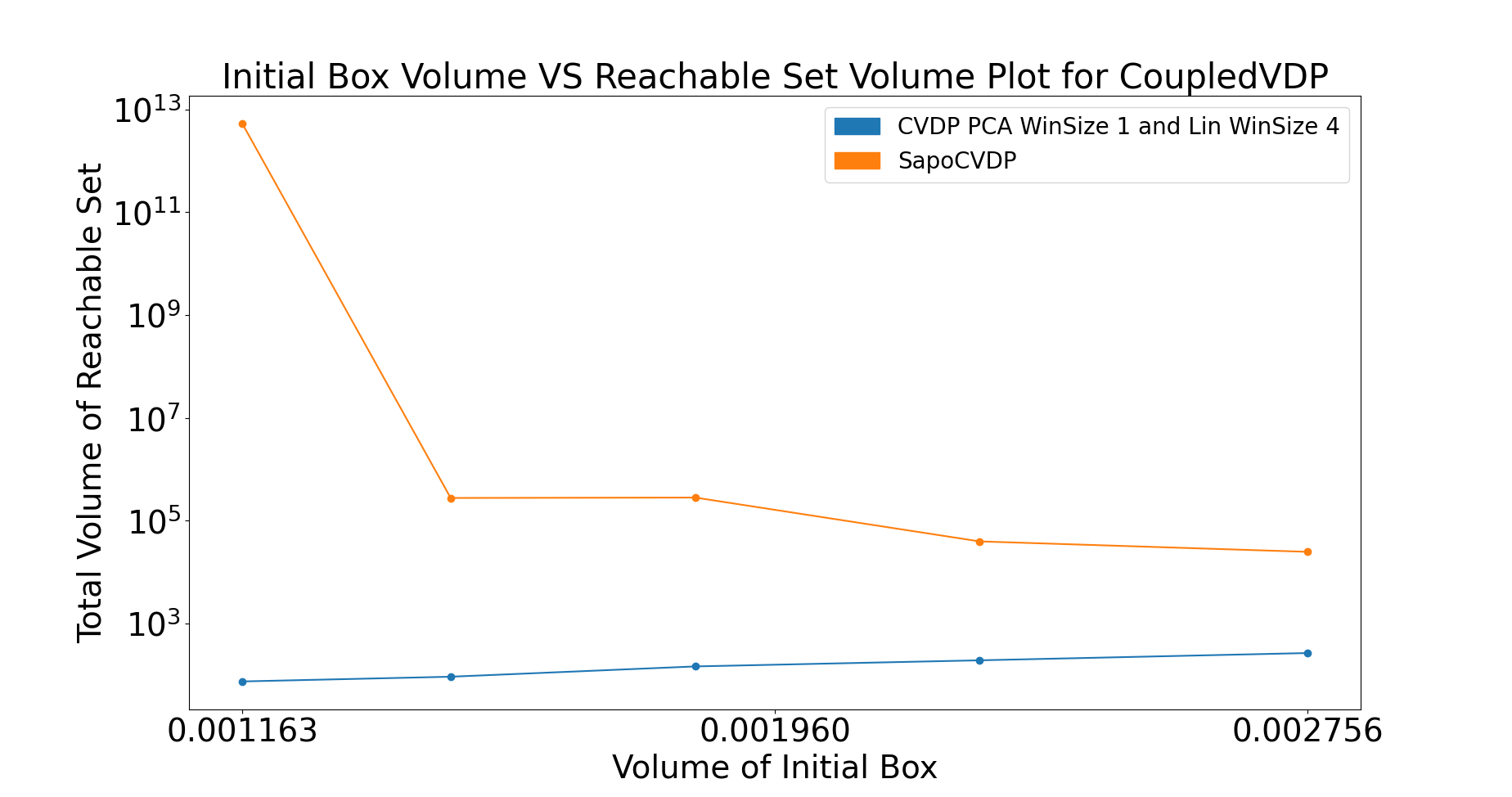}
    \caption{Coupled Vanderpol}
    \end{subfigure}%
    %\hspace{1em}
    \begin{subfigure}{0.5\textwidth}
    \centering
    \includegraphics[width=1.1\textwidth, height=0.75\textwidth]{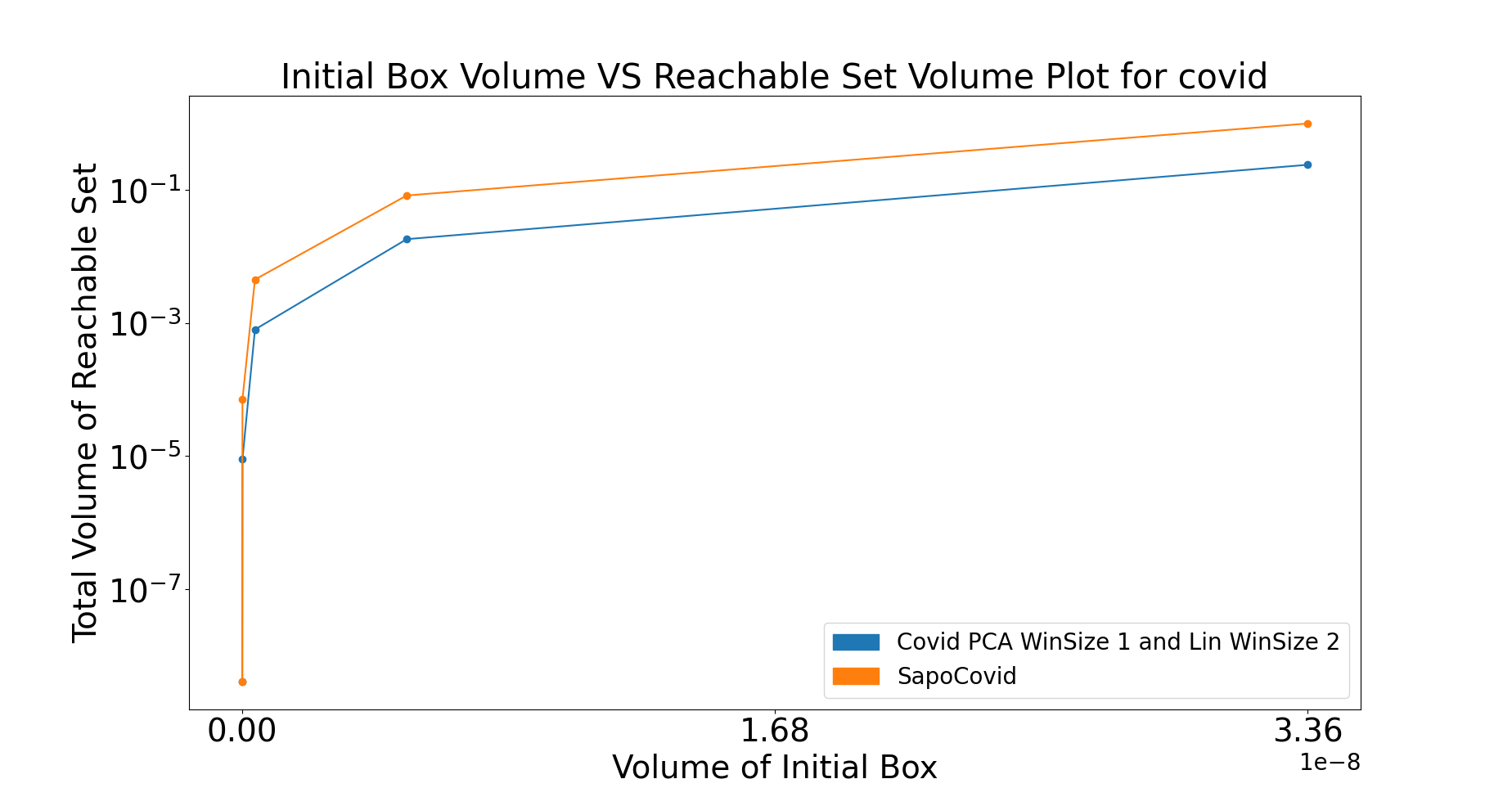}
    \caption{COVID}
    \end{subfigure}

    \caption{Comparison between the performance of diagonal static parallelotope bundles and that of the best performing dynamic parallelotope bundles as the volume of the initial set grows.}
    \label{fig:InitVolReachComp}
\end{figure}

A key advantage of our dynamic strategies is the improved ability to control the wrapping error naturally arising from larger initial sets. Figure \ref{fig:InitVolReachComp} presents charts showcasing the effect of increasing initial sets on the total flowpipe volume. We vary the initial box dimensions to gradually increase the box's volume. We then plot the total flowpipe volume after running the benchmark. The same initial boxes are also used in computations using Sapo's static parallelotopes. The number of parallelotopes defined by PCA and Linear Approximation directions were chosen based on best performance as seen in Table \ref{tab:voltable}. We remark that our dynamic strategies perform better than static ones in controlling the total flowpipe volume as the initial set becomes larger. On the other hand, the performance of static parallelotopes tends to degrade rapidly as we increase the volume of the initial box.

\vspace{-1em}
\subsubsection{Performance against Random Static Templates}
\begin{figure}[h!]
    \hspace{-1.5em}
    \begin{subfigure}{0.5\textwidth}
    \centering
    \includegraphics[width=1.1\textwidth, height=0.75\textwidth]{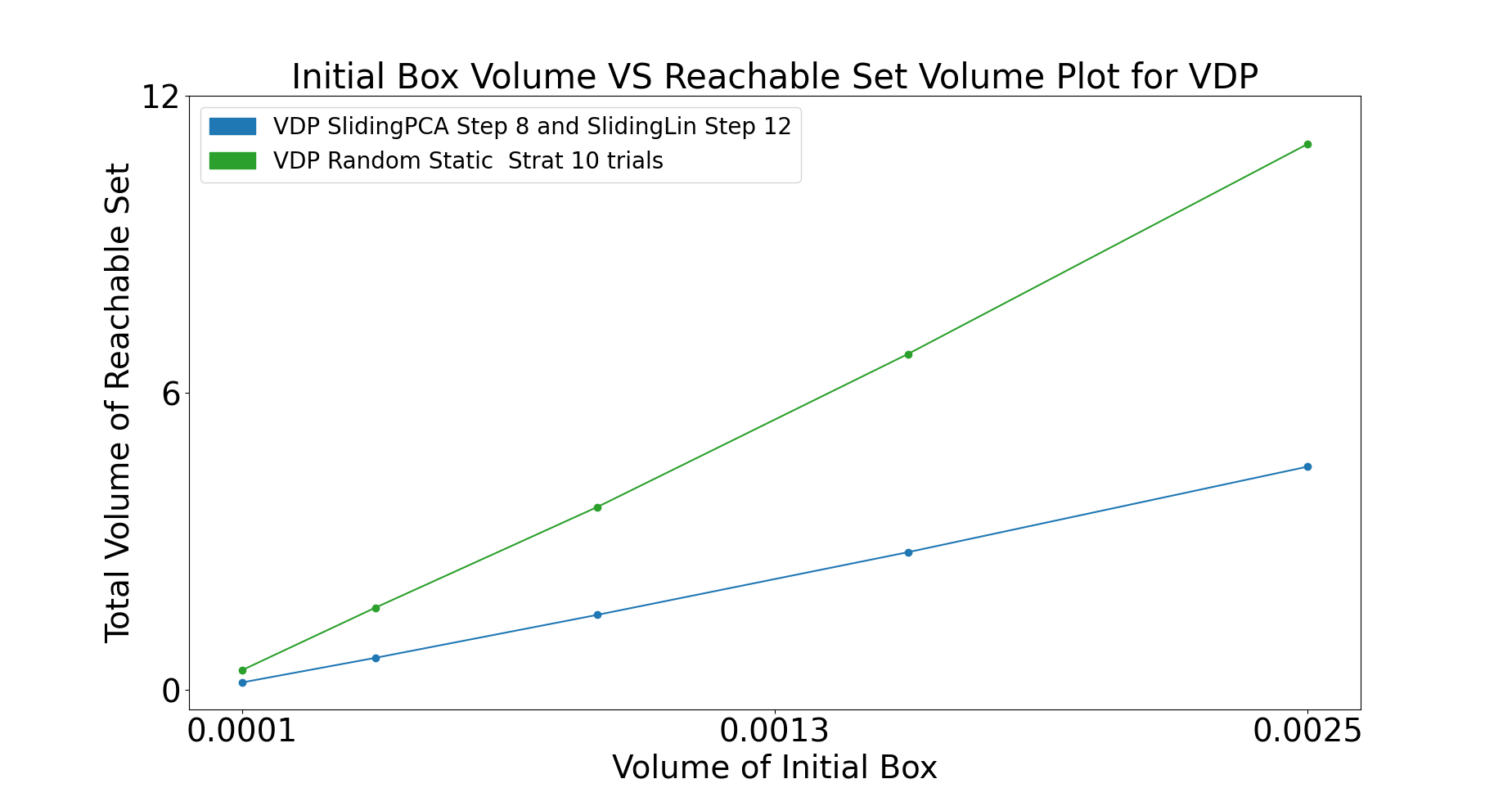}
    \caption{Vanderpol}
    \end{subfigure}%
    \begin{subfigure}{0.5\textwidth}
    \centering
    \includegraphics[width=1.1\textwidth, height=0.75\textwidth]{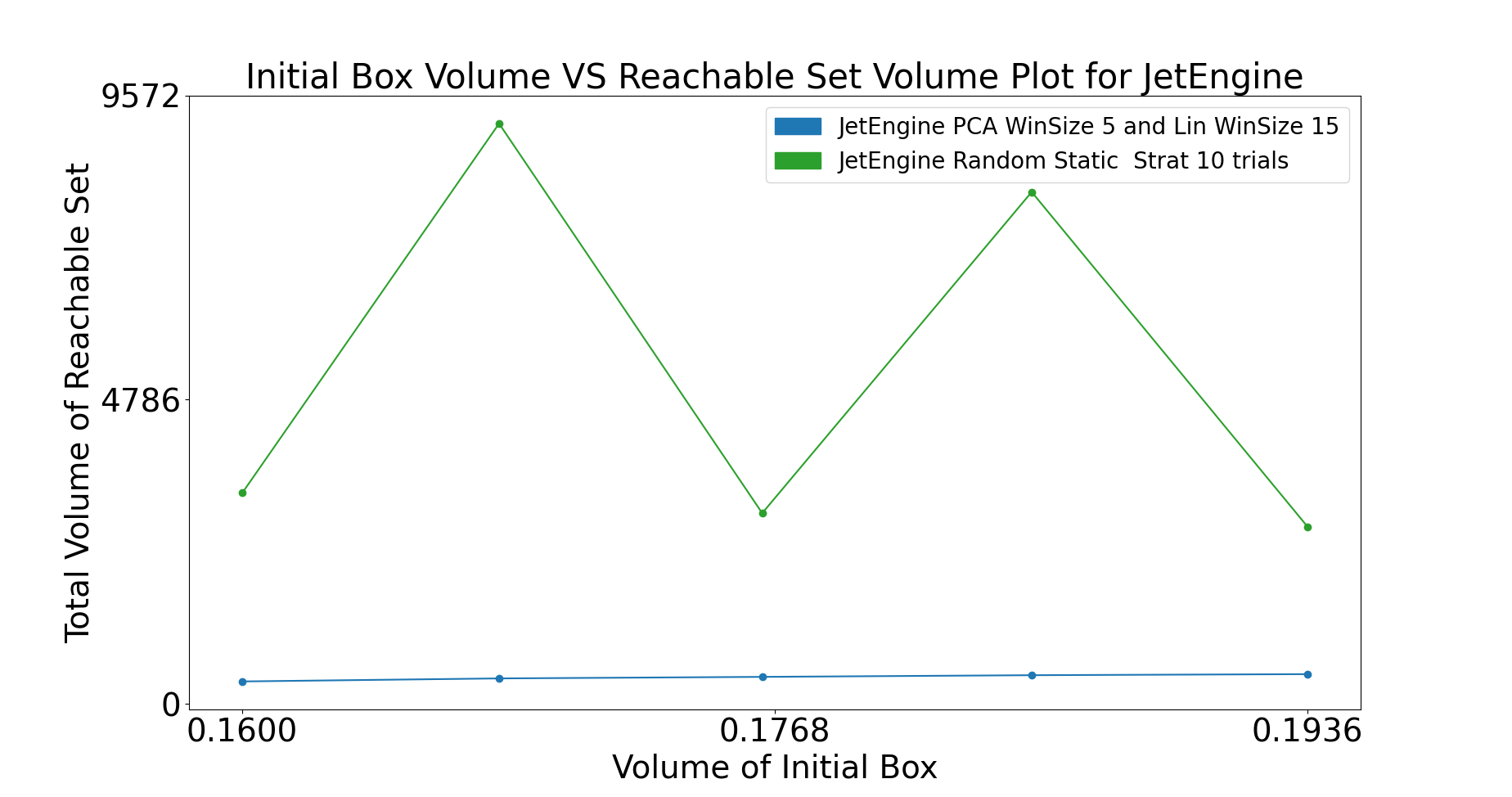}
    \caption{Jet Engine}
    \end{subfigure}

    \hspace{-1.5em}
    \begin{subfigure}{0.5\textwidth}
    \centering
    \includegraphics[width=1.1\textwidth, height=0.75\textwidth]{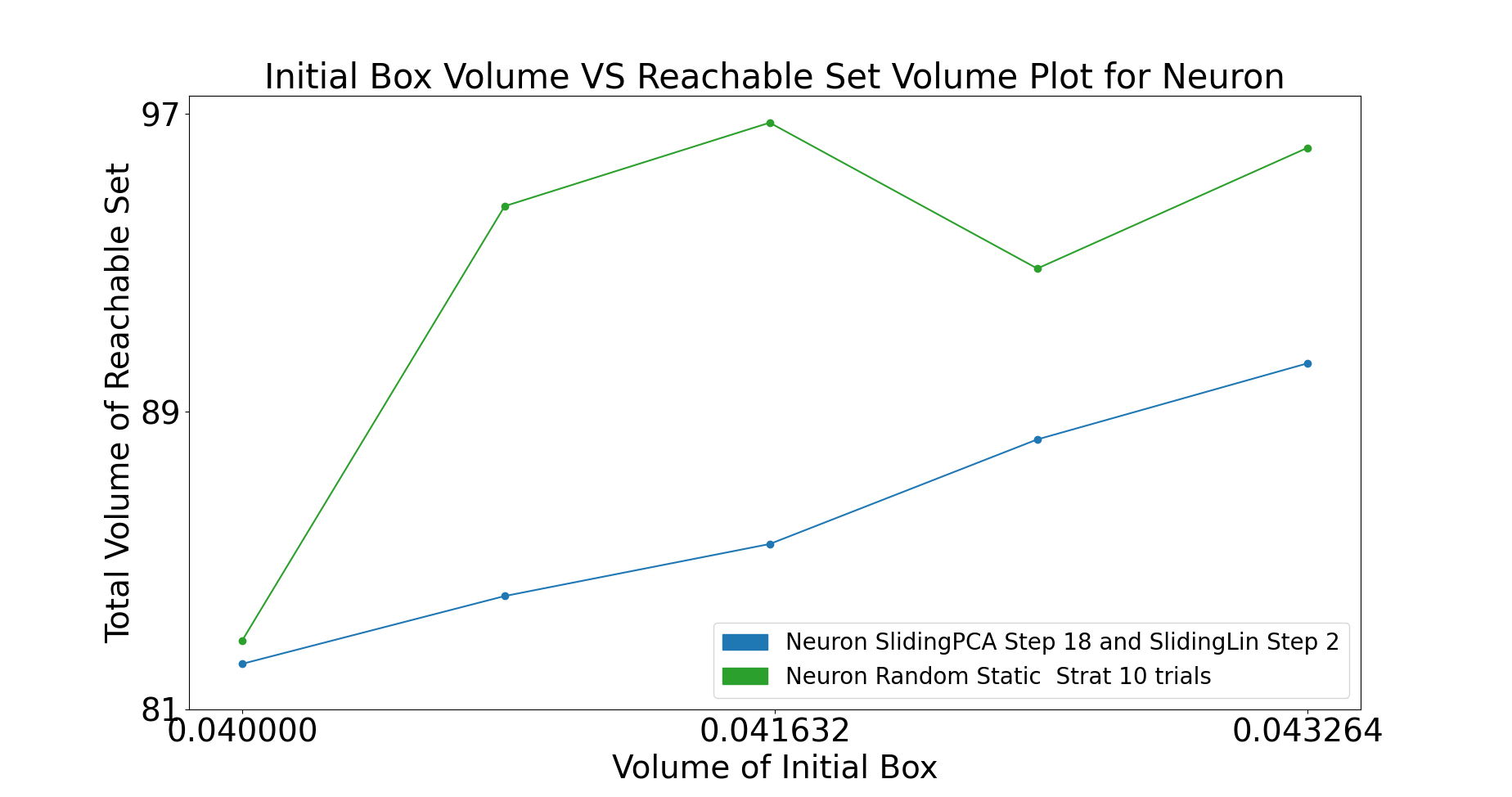}
    \caption{Neuron}
    \end{subfigure}%
    \begin{subfigure}{0.5\textwidth}
    \centering
    \includegraphics[width=1.1\textwidth, height=0.75\textwidth]{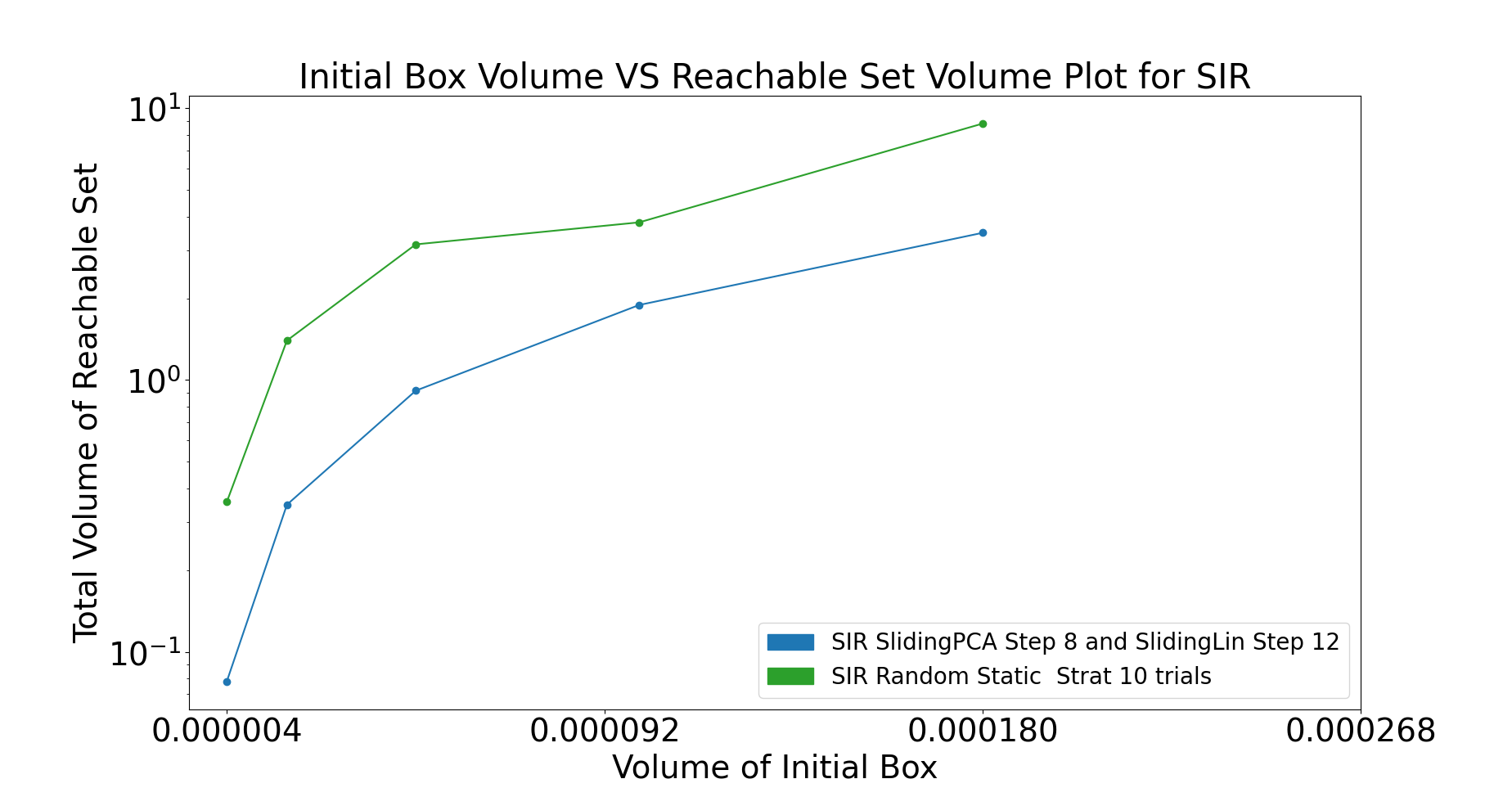}
    \caption{SIR}
    \end{subfigure}

    \hspace{-1.5em}
    \begin{subfigure}{0.5\textwidth}
    \centering
    \includegraphics[width=1.1\textwidth, height=0.75\textwidth]{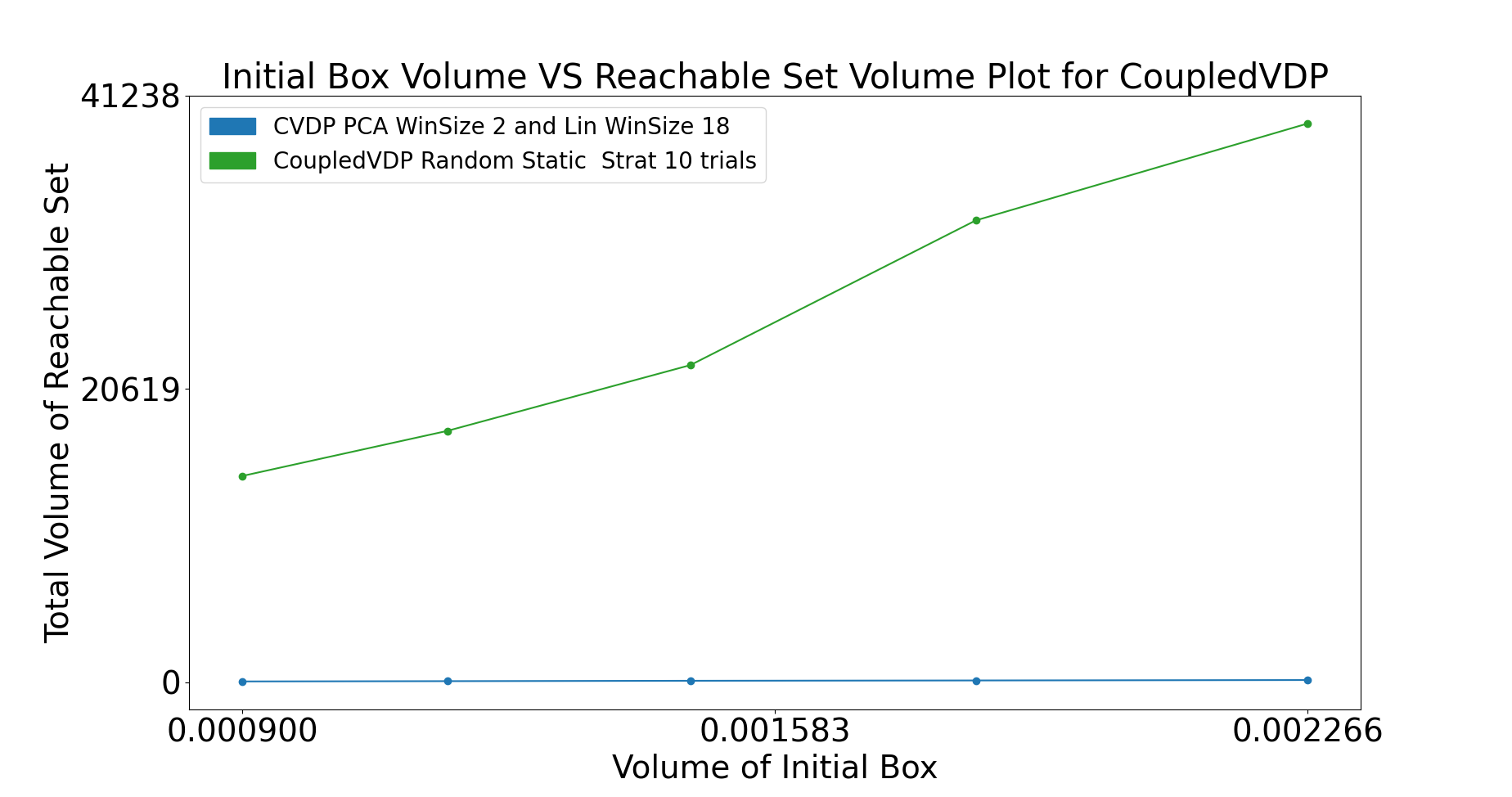}
    \caption{Coupled Vanderpol}
    \end{subfigure}%
    \begin{subfigure}{0.5\textwidth}
    \centering
    \includegraphics[width=1.1\textwidth, height=0.75\textwidth]{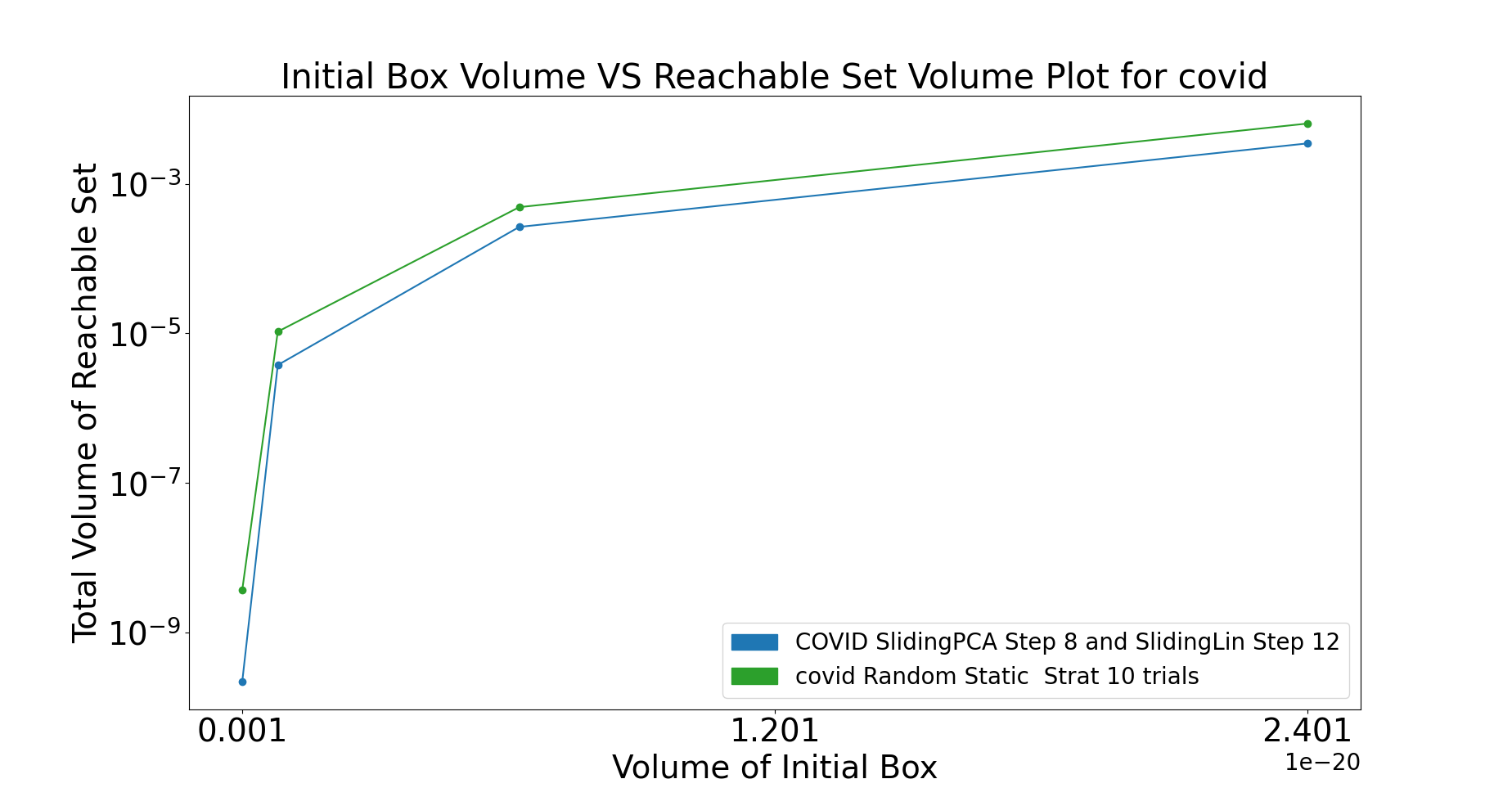}
    \caption{COVID}
    \end{subfigure}

    \caption{Comparision between random static strategies and the best performing dynamic strategies as the volume of the initial set grows. The total reachable set volumes for random static strategies are averaged over ten trials for each system.}
    \label{fig:RanStaticStratComp}
\end{figure}

We additionally benchmark our dynamic strategies against static random parallelotope bundles. We sample such parallelotopes in $n$ dimensions by first sampling a set of $n$ directions uniformly on the surface of the unit $(n-1)$-sphere, then defining our parallelotope using these sampled directions. We sample twenty of these parallelotopes for each trial and average the total flowpipe volumes. As shown in Figure \ref{fig:RanStaticStratComp}, our best-performing dynamic strategies consistently outperform static random strategies for all tested benchmarks.

\vspace{-0.8em}
\section{Conclusions}
In this paper, we investigated two techniques for generating templates dynamically: first using linear approximation of the dynamics, and second using PCA.
We demonstrated that these techniques improve the accuracy of reachable set by an order of magnitude when compared to static or random template directions.
We also observed that both these techniques improve the accuracy of the reachable sets for different benchamrks.
In future, we intend to investigate Koopman linearization techniques for computing alternative linear approximation template directions \cite{bak2021reachability}.
We also wish to investigate the use of a massively parallel implementation using HPC hardware such as GPUs for optimizing over an extremely large number of parallelotopes and their template directions. This is inspired by the approach behind the recent tool \emph{PIRK} \cite{devonport2020pirk}.

\subsection{Acknowledgements}
Parasara Sridhar Duggirala and Edward Kim acknowledge the support of the Air Force Office of Scientific Research under award number FA9550-19-1-0288 and FA9550-21-1-0121 and National Science Foundation (NSF) under grant numbers CNS 1935724 and CNS 2038960. Any opinions, findings, and conclusions or recommendations expressed in this material are those of the author(s) and do not necessarily reflect the views of the United States Air Force or the National Science Foundation.

\begin{table}[h!]
\hspace{1em}
\begin{subtable}[h]{0.45\textwidth}
     \centering
     \begin{tabular}{|c|c|}
     \hline
     Strategy & Total  Volume \\
     \hline\
     5 LinApp & 0.227911 \\
     \hline
     1 PCA, 4 LinApp& 0.225917 \\
     \hline
     2 PCA, 3 LinApp & 0.195573 \\
     \hline
     {\bf 3 PCA, 2 LinApp} & {\bf 0.188873} \\
     \hline
     4 PCA, 1 LinApp & 1.227753\\
     \hline
     5 PCA & 1.509897 \\
     \hline
     5 Static Diagonal(Sapo) & 2.863307  \\
     \hline
    \end{tabular}
    \caption{Vanderpol}
    \label{tab:vdpvol}
 \end{subtable}\hspace{1em}
 \begin{subtable}[h]{0.45\textwidth}
      \centering
      \begin{tabular}{|c|c|}
      \hline
      Strategy & Total  Volume \\
      \hline
      5 LinApp & 58199.62 \\
      \hline
      1 PCA, 4 LinApp & 31486.16 \\
      \hline
      {\bf 2 PCA, 3 LinApp} & {\bf 5204.09}\\
      \hline
      3 PCA, 2 LinApp & 6681.76 \\
      \hline
      4 PCA, 1 LinApp& 50505.10 \\
      \hline
      5 PCA  & 84191.15 \\
      \hline
      5 Static Diagonal (Sapo) & 66182.18  \\
      \hline
     \end{tabular}
     \caption{Jet Engine}
     \label{tab:enginevol}
  \end{subtable}

  \hspace{1em}
  \begin{subtable}[h]{0.45\textwidth}
       \centering
       \begin{tabular}{|c|c|}
       \hline
       Strategy & Total  Volume \\
       \hline
       5 LinApp  & 154.078\\
       \hline
       1 PCA, 4 LinApp  & 136.089\\
       \hline
       2 PCA, 3 LinApp  & 73.420\\
       \hline
       {\bf 3 PCA , 2 LinApp } & {\bf 73.126} \\
       \hline
       4 PCA, 1 LinApp  & 76.33 \\
       \hline
       5 PCA & 83.896 \\
       \hline
       5 Static Diagonal  (Sapo) & 202.406  \\
       \hline
      \end{tabular}
      \caption{FitzHugh-Nagumo}
      \end{subtable} \hspace{1em}
      \begin{subtable}[h]{0.45\textwidth}
        \centering
        \begin{tabular}{|c|c|}
        \hline
        Strategy & Total  Volume \\
        \hline
        {\bf 2 LinApp } & {\bf 0.001423} \\
        \hline
        1 PCA, 1 LinApp & 0.106546\\

        \hline
        2 PCA  & 0.117347\\
        \hline
        2 Static Diagonal (Sapo) & 0.020894\\
        \hline
       \end{tabular}
       \caption{SIR}
       \label{tab:sirvol}
    \end{subtable}

    \hspace{1em}
    \begin{subtable}[h]{0.45\textwidth}
         \centering
         \begin{tabular}{|c|c|}
         \hline
         Strategy & Total  Volume \\
         \hline
         5 LinApp & 5.5171 \\
         \hline
         {\bf 1 PCA, 4 LinApp } & {\bf 5.2536} \\
         \hline
         2 PCA, 3 LinApp  & 5.6670\\
         \hline
         3 PCA, 2 LinApp  & 5.5824\\
         \hline
         4 PCA, 1 LinApp  & 312.2108 \\
         \hline
         5 PCA  & 388.0513 \\
         \hline
         5 Static Diagonal (Best) & 3023.4463  \\
         \hline
        \end{tabular}
        \caption{Coupled Vanderpol}
        \label{tab:sirvol}
     \end{subtable}\hspace{1 em}
    \begin{subtable}[h]{0.45\textwidth}
         \centering
         \begin{tabular}{|c|c|}
         \hline
         Strategy & Total  Volume \\
         \hline
         3 LinApp & $2.95582227 * 10^{-10}$ \\
         \hline
         {\bf 1 PCA, 2 LinApp } & {\bf $2.33007583 * 10^{-10}$}\\
         \hline
         2 PCA, 1 LinApp &$ 4.02751770 * 10^{-9}$\\
         \hline
         3 PCA & $4.02749571 * 10^{-9}$\\
         \hline
         3 Static Diagonal (Sapo) & $4.02749571 * 10^{-9}$\\
         \hline
        \end{tabular}
        \caption{COVID}
        \label{tab:covidvol}
     \end{subtable}
 \caption{Tables presenting upper bounds on the total reachable set volume by strategy. The static directions are retrieved and/or inspired from Sapo models of equal dimension for benchmarking. The best performing strategy is highlighted in bold.}
     \label{tab:voltable}
\end{table}
\clearpage

\clearpage

\clearpage

\newpage
\nocite{*}
\bibliographystyle{abbrv}
\bibliography{templates}

\end{document}